\documentclass[reqno,11pt]{amsart}
\usepackage{fullpage}
\usepackage{amsfonts}
\usepackage{amssymb}
\usepackage{arydshln}
\usepackage{latexsym}
\usepackage{cite}
\usepackage{epic,graphicx}
\usepackage{color}
\usepackage{bm}
\usepackage{graphics,amsmath,amsthm,accents}

\numberwithin{equation}{section}
\numberwithin{equation}{section}

\newtheorem{Thm}{Theorem}

\newcommand{\nn}{\nonumber}
\newcommand{\wt}{\widetilde}
\newcommand{\wh}{\widehat}

\begin{document}

\title{Rational solutions for three semi-discrete modified Korteweg-de Vries type equations}

\author{
Ying-ying Sun$^1$, Song-lin Zhao$^{2*}$\\
\\\lowercase{\scshape{${}^1$
Department of Mathematics, University of Shanghai for Science and Technology, Shanghai,
200093, P.R. China}} \\
\lowercase{\scshape{
${}^2$ Department of Applied Mathematics, Zhejiang University of Technology,
Hangzhou 310023, P.R. China}}}
\email{*Corresponding Author: songlinzhao@zjut.edu.cn}

\begin{abstract}

In this paper, we consider three semi-discrete modified Korteweg-de Vries
type equations which are  the nonlinear lumped self-dual network equation,
the semi-discrete lattice potential modified Korteweg-de Vries equation and
a semi-discrete modified Korteweg-de Vries equation.
We derive several kinds of exact solutions, in particular rational solutions,
in terms of the Casorati determinant for these three equations respectively.
For some rational solutions, we present the related asymptotic analysis to understand their dynamics better.

\end{abstract}

\keywords{semi-discrete mKdV type equations, Casorati determinant solutions, rational solutions}

\maketitle

\section{Introduction}
\label{sec-1}

The study of rational solutions for integrable systems
is always a hot topic in mathematical physics and a source of contacts with other fields. For instance, the spatial-temporary localized
waves are used to describe rogue waves \cite{AAT}. Historically, the rational solutions of the
famous Korteweg-de Vries (KdV) equation was firstly investigated by Moser and his collaborators \cite{Moser,Moser-1}. Superficially,
the soliton solutions and rational solutions are
different with respect to their expressions, but in real
sense they are deeply related with each other. In general, rational solutions, expressed by a fraction of polynomials
of independent variables, can be derived from soliton solutions through a special limit procedure (see Refs. \cite{AS,Z-KdV} as examples).
Compared with the case in continuous integrable systems, it is more difficult to get rational solutions of the semi-discrete
integrable systems, or known as integrable differential-difference systems.
In spite of this, until now much progress has been made. Using the ``long wave limit''
procedure, C\^{a}rstea \textit{et al.} studied non-singular rational solutions for the 1-dimensional generalized
Volterra system \cite{Car} and Toda lattice \cite{Car-1}. Applying Hirota's bilinear
formalism and B\"{a}cklund transformation, Hu and Clarkson \cite{HC} gave the nonlinear superposition formulae of rational solutions for
three semi-discrete KdV type equations, including differential-difference KdV equation, Toda lattice and a semi-discrete KdV equation.
In \cite{WZ}, Wu and Zhang obtained rational solutions in the Casorati determinant form for the Toda lattice and the differential-difference KdV
equation.

Among those methods, Wronskian/Casoratian technique is one of powerful tools to construct rational solutions for the
integrable equations. Casoratian technique is usually regarded as the discrete analogue of Wronskian technique.
In general, in the construction of rational
solutions by using the Wronskian/Casoratian technique, the element vector should satisfy a linear differential/difference equation
set, in which the coefficient matrix always has only multiple zero eigenvalues
(please see our review paper \cite{Z-KdV} and references therein). Unfortunately, for the modified Korteweg-de Vries (mKdV) equation
no rational solutions arise from this perspective, which means such coefficient matrix is not allowed to
appear in the differential equation set. To eliminate this defect, a Galilean transformation was applied to transform
the mKdV equation into the KdV-mKdV equation, and the question about how to derive the rational
solutions of the former equation turned to those of the latter one\cite{SZ,ZZSZ}.

Motivated by the way constructing rational solutions in Refs. \cite{SZ,ZZSZ} and the relation between the mKdV equation and
semi-discrete mKdV type equations, in this paper we focus on the following three differential-difference equations
\begin{itemize}
\item{nonlinear lumped self-dual network equation (NLSNE)
\begin{align}
\label{NLSNE}
\partial_{tt} u_n
=\big(1+(\partial_t u_n)^2\big)\big(\tan(u_{n-1}-u_n)-\tan(u_n-u_{n+1})\big),
\end{align}}
\item{semi-discrete lattice potential mKdV (sd-lpmKdV) equation
\begin{align}
\label{sdpmKdV}
p \partial_t \ln v_{n}=\frac{v_{n-1}-v_{n+1}}{v_{n-1}+v_{n+1}},
\end{align}}
\item{semi-discrete mKdV (sd-mKdV) equation
\begin{align}
\label{sdmKdV}
\partial_t w_n=(1+w_n^2)(w_{n-1}-w_{n+1}),
\end{align}}
\end{itemize}
and aim to find their exact solutions expressed by Casorati determinants. The dependent variables $u$,
$v$ and $w$ adopted in \eqref{NLSNE}-\eqref{sdmKdV} are
defined on the discrete-continuous coordinates $(n,t)\in \mathbb{Z} \times \mathbb{R}$.
The parameter $p$ appeared in \eqref{sdpmKdV} is
the continuous lattice parameter associated with the grid size in the direction $n$.
In weakly nonlinear and continuum limits, both the NLSNE and sd-lpmKdV equation reduce to the
continuous pmKdV equation, while the sd-mKdV equation reduces to the
continuous mKdV equation. Thus these three equations are viewed as semi-discrete mKdV type equations.
The semi-discrete mKdV type equations arise in a wide variety of fields,
such as plasma physics, electromagnetic waves
in ferromagnetic, antiferromagnetic or dielectric systems.
The NLSNE \eqref{NLSNE}, also named Hirota lattice equation  \cite{BM}, was initialed to
describe the transmission line with the nonlinear capacitances and inductances \cite{H-NLSNE}.
This equation was also used to model one-dimensional anharmonic chain of
atoms, where $u_n$ is the displacement of the $n$-th atom from the equilibrium position in the lattice.
The sd-lpmKdV equation \eqref{sdpmKdV} was obtained from the fully discrete
lpmKdV equation by applying the skew continuum limit \cite{Sch-eq}. Whereas the fully discrete lpmKdV equation has
multidimensional consistency, the equation \eqref{sdpmKdV} should also have this property
in some sense. The sd-mKdV equation \eqref{sdmKdV}, also known as the modified Volterra equation \cite{sdmkdv-1976},
was firstly proposed by Ablowitz and Ladik \cite{Ab-1977} when they
studied the conceptual analogy between Fourier analysis and the exact solutions for a
class of nonlinear differential-difference equations.

With regard to solutions of the equations \eqref{NLSNE}-\eqref{sdmKdV},
some results have been obtained. It has been shown that
equation \eqref{NLSNE} possesses multi-soliton solutions in terms of sums of
exponentials \cite{H-NLSNE} and Casorati determinant form \cite{Nimmo}. Except for solitons,
breather solutions, limit breather solutions, wobbling kink solutions and nonlinear periodic waves
solutions were also derived \cite{ZZZ,Lap1,Lap2}.
Recently, soliton solutions and Jordan-block solutions for the equation
\eqref{sdpmKdV} \cite{Zhao-dpAKNS} was derived through the generalized Cauchy matrix approach \cite{ZZ}. For the
sd-mKdV equation \eqref{sdmKdV}, many approaches, such as inverse scattering transform \cite{AS-1981,LDZ}, Darboux transformation \cite{Gao-YT,Li},
bilinear approach \cite{ZJD}, discrete Jacobi sub-equation method \cite{WM}, algebro-geometric approach \cite{GG},
Riemann-Hilbert approach \cite{ZGK} and Deift-Zhou nonlinear steepest descent method \cite{CF}, have been developed
to construct its exact solutions.

In this paper, we mainly study the rational solutions for the semi-discrete mKdV type equations \eqref{NLSNE}-\eqref{sdmKdV}.
The treatment for these three equations follows
closely from the derivations of rational solutions of the mKdV equation in \cite{SZ,ZZSZ}.
The paper is organized as follows. In Sec. 2, we derive several kinds of exact solutions for the NLSNE \eqref{NLSNE},
in particular the non-singular rational solutions. Moreover,
dynamics for the first two non-trivial rational solutions are analyzed by using asymptotic analysis.
In Sec. 3 and Sec. 4, the rational solutions and dynamics for sd-lpmKdV equation
\eqref{sdpmKdV} and sd-mKdV equation \eqref{sdmKdV} are shown respectively.
Sec. 5 is devoted to the conclusions. In addition, an appendix is given as complement
to the paper.

\section{Rational solutions for the NLSNE \eqref{NLSNE}}

The purpose of this section is to find the rational solutions for the NLSNE \eqref{NLSNE}.
We first introduce a Galilean transformation, with which an equivalent form of
the NLSNE \eqref{NLSNE} is derived. We then solve the equivalent equation through
bilinear B\"{a}cklund transformation and obtain soliton solutions, Jordan-block solutions
and rational solutions. Dynamics of some rational solutions are discussed as well.

\subsection{Bilinear B\"{a}cklund transformation and Casorati determinant solutions}

Inspired by the idea how to construct non-singular rational solutions for the mKdV equation \cite{ZZSZ},
in this section, we start by introducing a Galilean transformation
\begin{align}
\label{z-t}
u(n,t)=U(n,z), \quad z=\frac{2}{\sqrt{4-c^2}}t,\quad c\in (-2,2)
\end{align}
for the NLSNE \eqref{NLSNE}. Then the NLSNE \eqref{NLSNE} can be rearranged as
\begin{align}
\label{NLSNE-c}
\partial_z^2 U_n=\left(1-\frac{c^2}{4}+(\partial_z U_n)^2\right)\left(\tan(U_{n-1}-U_n)-\tan(U_n-U_{n+1})\right).
\end{align}
By a dependent variable transformation
\begin{align}
\label{U-Tran}
U_n=\frac{i}{2}\ln\frac{f^*_n}{f_n}+\frac{c}{2}z,
\end{align}
the equation \eqref{NLSNE-c} can be transformed into the bilinear equations (cf. \cite{H-NLSNE,Nimmo})
\begin{subequations}
\label{bili}
\begin{align}
& \label{bili-a}\left(D^2_z-4\sinh^2\frac{D_n}{2}\right)f_n\cdot f_n=0, \\
& \label{bili-b}\left(D^2_z+2icD_z+4\sinh^2\frac{D_n}{2}\right)f_n\cdot f^*_n=0,
\end{align}
\end{subequations}
where $i$ is the imaginary unit, $f_n\doteq f(n,z)$, asterisk stands for the complex conjugate,
and the Hirota's bilinear operator $D^m_tD^n_x $ and bilinear difference operator $e^{\epsilon D_n}$ are defined by \cite{Hi-Bi}
\begin{subequations}
\label{Hiro-ope}
\begin{align}
& D^m_tD^n_x a(t,x)\cdot b(t,x)=\frac{\partial^m}{\partial s^m}
\frac{\partial^n}{\partial y^n} a(t+s,x+y)b(t-s,x-y)\big|_{s=0,y=0}, \\
& e^{\epsilon D_n}f_n\cdot g_n=f_{n+\epsilon}g_{n-\epsilon}.
\end{align}
\end{subequations}
Using the classical method given by Hirota in Ref.\cite{Hirota-BT-PTP}, we derive the following bilinear
B\"{a}cklund transformation linking two solutions $f_n,~g_n$ of the bilinear equations \eqref{bili} 
\begin{subequations}
\label{BT-bili}
\begin{align}
& \label{BT-bili-a} (D_z+e^{-D_n}-\lambda_1)f_n\cdot g_n=0, \\
& \label{BT-bili-b} (D_z-e^{D_n})f_n\cdot g_{n+1}=0, \\
& \label{BT-bili-c} (D_z-e^{-D_n}+ic)f_n\cdot g^*_{n}+i\lambda_2 f^*_n\cdot g_{n}=0,
\end{align}
\end{subequations}
where $\lambda_1, \lambda_2 \in \mathbb{C}$ in \eqref{BT-bili} are two B\"{a}cklund parameters.

When $c=0$, Zhang \textit{et al.} \cite{ZZZ} constructed Casorati determinants
\begin{subequations}
\label{Caso}
\begin{align}
& f_n = \left|\begin{array}{cccc}
\phi_1(n) &  \phi_1(n+1)    & \cdots &   \phi_1(n+N-1) \\
\phi_2(n) &  \phi_2(n+1)    & \cdots &   \phi_2(n+N-1) \\
\vdots &  \vdots    & \cdots &   \vdots \\
\phi_N(n) &  \phi_N(n+1)    & \cdots &   \phi_N(n+N-1)
\end{array}\right|=|0,1,2,\ldots,N-1|=|\wh{N-1}|, \\
& g_n = \left|\begin{array}{cccc}
\phi_1(n) &      \cdots &   \phi_1(n+N-2)& 0 \\
\phi_2(n) &     \cdots &   \phi_2(n+N-2)& 0\\
\vdots &  \cdots& \vdots     &   \vdots \\
\phi_N(n) &   \cdots &   \phi_N(n+N-2) &1
\end{array}\right|=|0,1,\ldots,N-2,\tau_N|=|\wh{N-2},\tau_N|,
\end{align}
\end{subequations}
for the bilinear B\"{a}cklund transformation \eqref{BT-bili}, where
\begin{align}
\phi_j(n)=ia^+_je^{k_j n+e^{k_j}z}+a^-_je^{-k_j n+e^{-k_j}z},\quad a^{\pm}_j, k_j \in \mathbb{R}.
\end{align}
Here and hereafter we follow Freeman-Nimmo's notation \cite{Fr-Ni}; let $\wh{N-j}$ indicate the
set of consecutive columns $0, 1, 2, \ldots, N-j$. Besides, we introduce $\wt{N-j}$ to indicate the
set of columns $0, 2, 4 \ldots, 2(N-j)$.

Exact Casorati determinant solutions of \eqref{BT-bili} are presented by the following Theorem.
\begin{Thm}
\label{Thm-CES}
The Casorati determinants
\begin{align}
\label{fg-phi}
f_n=|\wh{N-1}|,\quad g_n=|\wh{N-2},\tau_N|,
\end{align}
solve the bilinear B\"{a}cklund transformation \eqref{BT-bili},
provided the basic column vectors $\phi(n)\doteq \phi(n,z)=(\phi_1(n),\phi_2(n),\cdots,\phi_N(n))^T$ satisfy the following condition equation set (CES)
\begin{subequations}
\label{CES}
\begin{align}
& \phi(n+1)+\phi(n-1)=A\phi(n), \\
& i(\phi(n+1)-\phi(n-1))-c\phi(n)=B\phi^*(n), \\
& \partial_z \phi(n)=\phi(n+1).
\end{align}
\end{subequations}
Here,
\begin{align}
A=\left(
\begin{array}{cc}
\wt{A} &  \bm 0  \\
*_1 &\lambda_1
\end{array}\right), \quad B=\left(
\begin{array}{cc}
\wt{B} &  \bm 0  \\
*_2 &\lambda_2
\end{array}\right)
\end{align}
are two arbitrary $N\times N$ invertible matrices and satisfy
\begin{align}
\label{AB}
BB^*=A^2-(4-c^2)I,
\end{align}
where $\wt{A},~\wt{B}$ are two $(N-1)\times (N-1)$ complex matrices, $\{*_1\}$ and $\{*_2\}$ are two
$(N-1)$-th order row vectors, $\bm 0$ stands for an $(N-1)$-th
order column zero vector, and $I$ is an  identity  matrix of size $N$.
\end{Thm}

Since this theorem can be proved similarly by the method given in \cite{ZZZ} and here we skip it.
From the expression \eqref{Caso} of $f_n$ and $g_n$ , we see that these two variables correspond to different sizes of solutions for \eqref{NLSNE-c}.
For example, $f_n$ and $g_n$ could serve for $N$- and $(N-1)$-soliton solutions through
\eqref{U-Tran} and $U_n=\frac{i}{2}\ln\frac{g^*_n}{g_n}+\frac{c}{2}z$, respectively. To construct the rational solutions for the NLSNE \eqref{NLSNE}
based on the Theorem \ref{Thm-CES}, in the following we assume that $c$ is a non-zero constant.

\subsection{Solutions to CES \eqref{CES}}

In this part we solve the CES \eqref{CES} and
give explicit solutions of the equation \eqref{sdmKdV-c}. Solutions
for the system \eqref{CES} can be classified according to the
canonical forms of $A$ and $B$. Let us list them out case by case.

\subsubsection{Soliton solutions}

Let $A$ be a diagonal matrix given by
\begin{subequations} \label{AB1}
\begin{align} \label{A1}
A=\mbox{diag}\left(2\cosh k_1,~2\cosh k_2,~\ldots,~2\cosh k_N\right)
\end{align}
with distinct non-zero real eigenvalues. Then from the relation between $A$ and $B$ \eqref{AB} we get
\begin{align} \label{B1}
B=\mbox{diag}\left(-\sqrt{c^2+4\sinh^2k_1},~-\sqrt{c^2+4\sinh^2k_2},~\ldots,~-\sqrt{c^2+4\sinh^2k_N}\right).
\end{align}
\end{subequations}
Substituting \eqref{AB1} for $A$ and $B$ into the CES \eqref{CES}, we obtain the explicit form of the Casorati elements
\begin{align}
\phi_j(n)=\sqrt{c+2i\sinh k_j}e^{k_j n+e^{k_j}z+\xi_j^{(0)}}+\sqrt{c-2i\sinh k_j}e^{-k_j n+e^{-k_j}z-\xi_j^{(0)}},
\end{align}
where $k_j, \xi_j^{(0)} \in \mathbb{R}, ~(j=1,\cdots,N)$. In this case we derive $N$-soliton solutions for the NLSNE \eqref{NLSNE}.

\subsubsection{Jordan-block solutions}

To present elements of the basic Casoratian column vector in this case,
we first introduce lower triangular Toeplitz (LTT) matrices which are defined as
\begin{equation*}
\mathcal{A}=\left(\begin{array}{cccccc}
a_0 & 0    & 0   & \cdots & 0   & 0 \\
a_1 & a_0  & 0   & \cdots & 0   & 0 \\
a_2 & a_1  & a_0 & \cdots & 0   & 0 \\
\vdots &\vdots &\cdots &\vdots &\vdots &\vdots \\
a_{N-1} & a_{N-2} & a_{N-3}  & \cdots &  a_1   & a_0
\end{array}\right)_{N\times N},~~~a_j\in \mathbb{C}.
\end{equation*}
Note that all the LTT matrices of same order compose a commutative set in terms of matrix product.
Canonical form of such a matrix is a Jordan matrix.
LTT matrices play an important role in generating multiple-pole (or limit) solutions (cf.\cite{Z-KdV,ZZSZ}).

Let $A$ and $B$ be LTT matrices
\begin{subequations}
\label{AB-J1}
\begin{align}
& A=(\alpha_{s,j})_{N\times N},\quad \alpha_{s,j}=\Biggl\{
\begin{array}{ll}
\frac{2}{(s-j)!}\partial^{s-j}_k\cosh k,&~s\geq j,\\
0,&~s<j,
\end{array}\\
& B=(\beta_{s,j})_{N\times N},\quad \beta_{s,j}=\Biggl\{
\begin{array}{ll}
\frac{-1}{(s-j)!}\partial^{s-j}_k\sqrt{c^2+4\sinh^2k},&~s\geq j,\\
0,&~s<j,
\end{array}
\end{align}
\end{subequations}
which satisfy the relation \eqref{AB}.
Then the vector $\phi(n)$ which is used to generate the Casorati determinant can be taken as
\begin{subequations}
\label{Ph-lim-Ph}
\begin{align}
\label{Ph-lim}
\phi(n)&=\mathcal{A}^+\phi^+(n)+\mathcal{A}^-\phi^-(n)
\end{align}
with
\begin{align}
\label{Ph-i}
\phi^{\pm}(n)&=(\phi^{\pm}_0(n),\phi^{\pm}_1(n),\ldots,\phi^{\pm}_{N-1}(n))^T, \\
\label{phi-k}
\phi^\pm_s(n)&=\frac{1}{s!}\partial_{k}^{s}[\sqrt{c\pm 2i\sinh k}e^{\pm k n+e^{\pm k}z+\pm \xi^{(0)}}],
\end{align}
\end{subequations}
where $\xi^{(0)} \in \mathbb{R}$ and $\mathcal{A}^{\pm}$ are two arbitrary LTT matrices of $N$-th order.

\subsubsection{Rational solutions}
\label{subsub-rs-NLSNE}

Rational solutions are formally generated from Jordan-block solutions by taking $k=0$ in \eqref{Ph-lim}
and \eqref{AB-J1}. However, to avoid trivial solutions, in practice we do the following.
Consider the generating function
\begin{align}
\label{phi-ra}
\rho(n)=\phi^+_0(n)+\phi^-_0(n)=\sqrt{c+2i\sinh k}e^{k n+e^{k}z}+\sqrt{c-2i\sinh k}e^{-kn+e^{-k}z},
\end{align}
where we have taken $\xi^{(0)}=0$ to guarantee this is an even function for $k$. The Taylor
expansion of $\rho(n)$ with respect to $k$ at $k=0$ can be written as
\begin{align}
\rho(n)=\sum_{j=1}^{+\infty}\psi_{j+1}k^{2j},
\end{align}
where
\begin{align}
\psi_{j+1}=\frac{1}{(2j)!}\frac{\partial^{2j}}{\partial k^{2j}}\rho(n)\bigg|_{k=0}, \quad (j= 0, 1, \ldots).
\end{align}
Then we obtain the vectors $\psi=(\psi_1,\psi_2,\ldots,\psi_N)^T$ satisfy the relations \eqref{CES} with
\begin{subequations}
\label{AB-J1-0}
\begin{align}
& A=(\alpha^0_{s,j})_{N\times N},\quad \alpha^0_{s,j}=\Biggl\{
\begin{array}{ll}
\frac{2}{(2(s-j))!},&~s\geq j,\\
0,&~s<j,
\end{array}\\
& B=(\beta^0_{s,j})_{N\times N},\quad \beta^0_{s,j}=\Biggl\{
\begin{array}{ll}
\frac{-1}{(2(s-j))!}\partial^{2(s-j)}_k\sqrt{c^2+4\sinh^2k}\bigg|_{k=0},&~s\geq j,\\
0,&~s<j.
\end{array}
\end{align}
\end{subequations}
Moreover, $f_n=|\wh{N-1}|$ and $g_n=|\wh{N-2},\tau_N|$ composed by $\psi$ are solutions of the bilinear B\"{a}cklund transformation
\eqref{BT-bili} with $\lambda_1=2$ and $\lambda_2=c$.
The first few explicit forms of some $f=\mbox{Cas}(\psi)$ are
\begin{subequations}
\begin{align}
& f_{N=1}=2 e^z \sqrt{c}, \label{fn=1} \\
& f_{N=2}=2 e^{2z}\big(c(1+2z+2n)+2i\big),  \label{fn=2}  \\
& f_{N=3}=-\frac{2e^{3z}}{3c^{\frac{3}{2}}}\bigg((1+z+n)\big(12c+(1-4(1+z+n)^2)c^3\big)+2zc^3 \nn \\
&~~\qquad\quad-12i\big(1-1/4c^2+(1+z+n)^2c^2\big)\bigg).  \label{fn=3}
\end{align}
\end{subequations}
Rational solutions of NLSNE \eqref{NLSNE} follows immediately from these solutions by \eqref{z-t} and \eqref{U-Tran}.

\subsection{Rational solutions to NLSNE \eqref{NLSNE} and dynamics}
\label{sec-212}

By the transformations \eqref{z-t} and \eqref{U-Tran} and noting that
\begin{align}
\frac{i}{2}\ln\frac{f^*_n}{f_n}=\arctan \frac{\mbox{Im}f_n}{\mbox{Re}f_n}=\frac{\pi}{2}-\arctan \frac{\mbox{Re}f_n}{\mbox{Im}f_n},
\end{align}
the rational solutions of NLSNE \eqref{NLSNE} can be expressed as
\begin{align}
u=\frac{\pi}{2}+\frac{c}{2}z-\arctan \frac{\mbox{Re}f_n}{\mbox{Im}f_n},\quad z=\frac{2}{\sqrt{4-c^2}}t.
\end{align}
The first two non-trivial rational solutions of \eqref{NLSNE} read
\begin{subequations} \label{rational-NLSNE-2}
\begin{align}
& \label{solu-1} u=\frac{\pi}{2}+\frac{c}{2}z-\arctan\big(c(\frac{1}{2}+z+n)\big), \\
& \label{solu-2} u=\frac{\pi}{2}+\frac{c}{2}z+\arctan\bigg(\frac{8(1+z+n)c+zc^3}
{6\big(1-1/4c^2+(1+z+n)^2c^2\big)}-\frac{(1+z+n)c}{3}\bigg)
\end{align}
\end{subequations}
with $z=\frac{2}{\sqrt{4-c^2}}t$,
which are corresponding to \eqref{fn=2} and \eqref{fn=3} respectively. Note that \eqref{fn=1} leads to a trivial solution $u=\frac{c}{\sqrt{4-c^2}}t$.

The solution \eqref{solu-1} is a non-singular traveling wave.
For a fixed $t$, $u$ appears as kink shape $(c<0)$ or anti-kink shape $(c>0)$ with wave span $\pi$,
as depicted in Fig. 1. One can identify the solution \eqref{solu-1} by its moving inflexion point trace
\[n(t)=-\frac{2}{\sqrt{4-c^2}}t-\frac{1}{2},\]
the slope  $c$, and the value of at the inflexion point $\frac{\pi}{2}+\frac{c}{\sqrt{4-c^2}}t$, respectively.


\begin{center}
\begin{picture}(120,90)
\put(-100,-23){\resizebox{!}{3cm}{\includegraphics{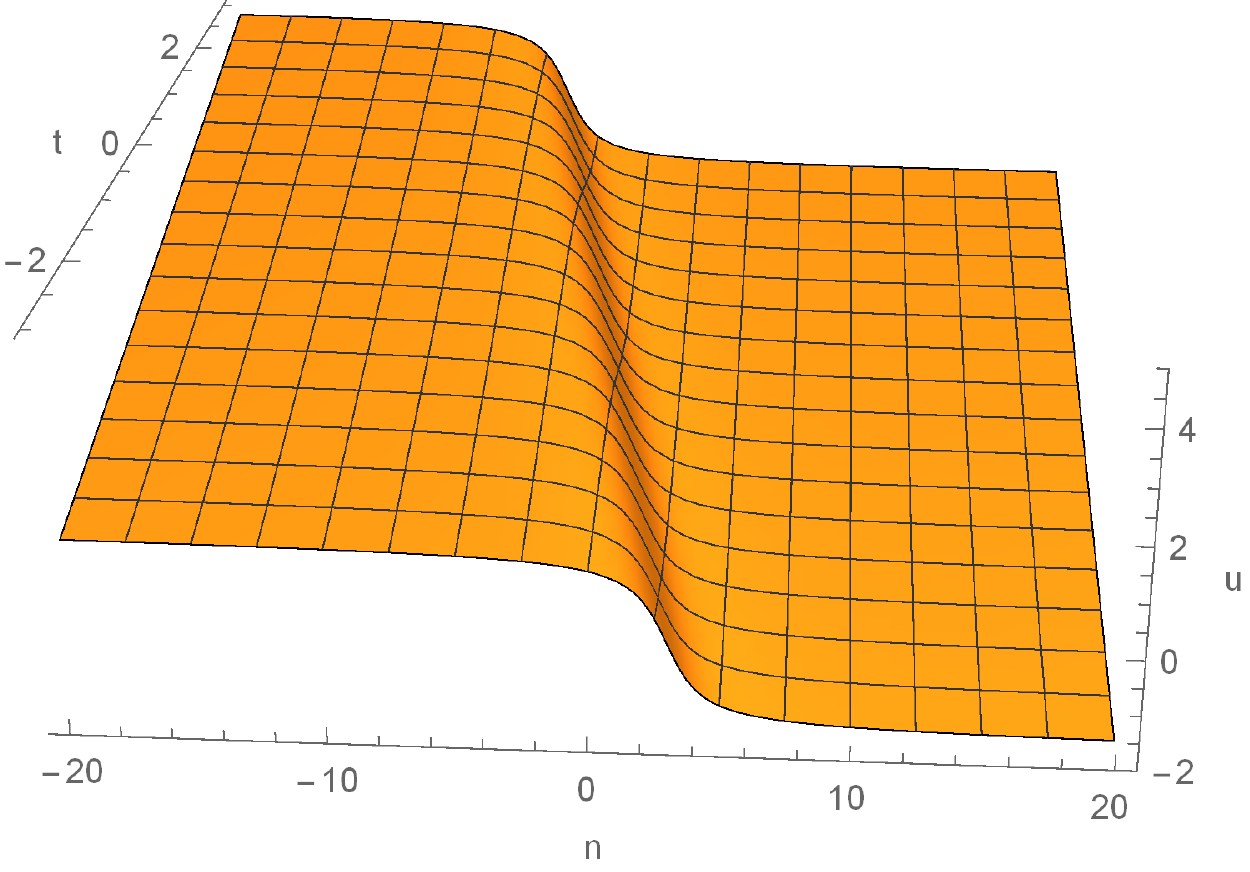}}}
\put(70,-23){\resizebox{!}{4cm}{\includegraphics{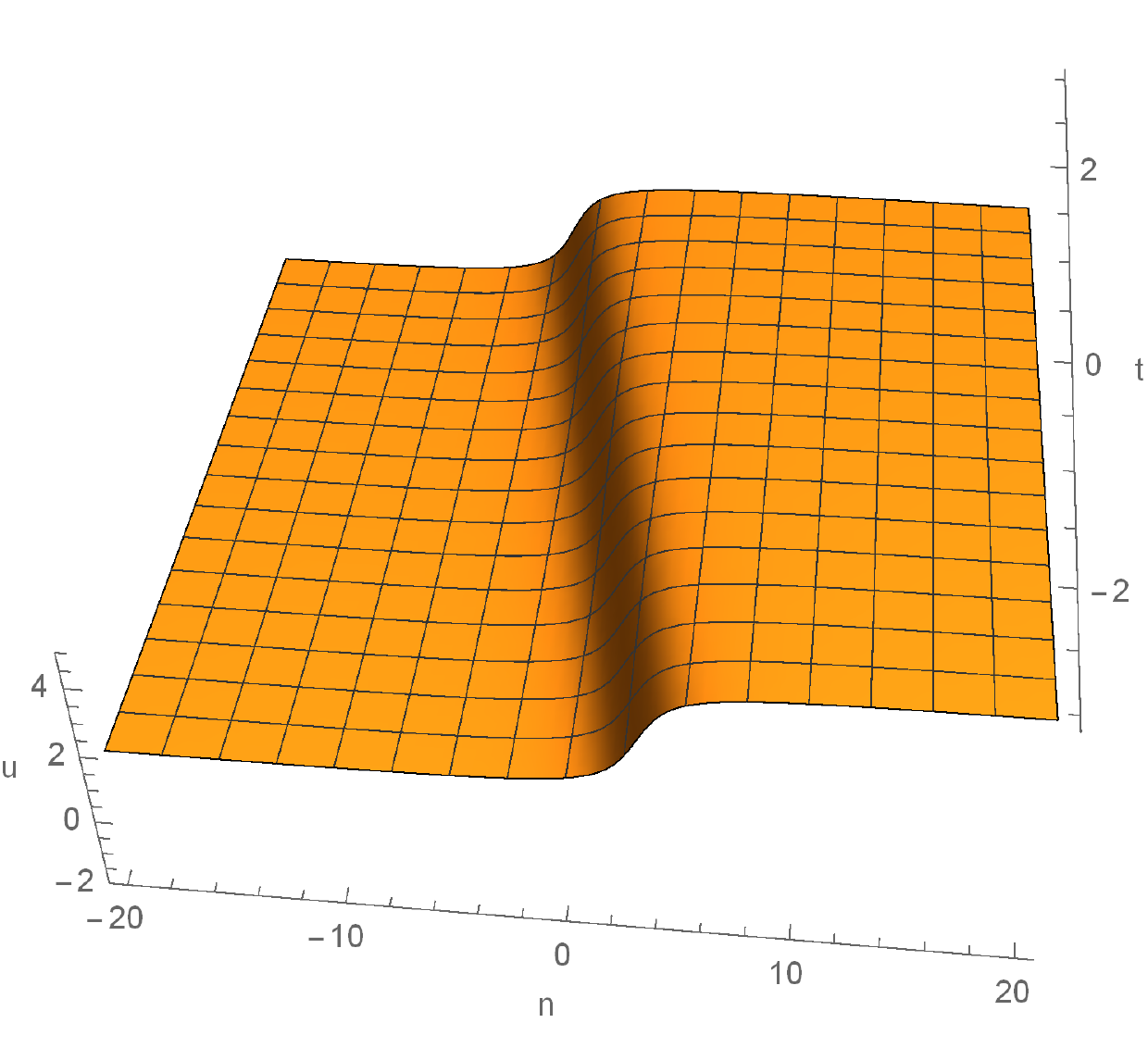}}}
\end{picture}
\end{center}
\vskip 20pt
\begin{center}
\begin{minipage}{11cm}{\footnotesize
\qquad\qquad\qquad(a)\qquad\qquad\qquad\qquad\qquad\qquad\qquad\qquad\qquad (b)\\
{\bf Fig. 1} The shape and movement of the solution \eqref{solu-1}. (a) Solution for $c=1$; (b)
Solution for $c=-1$.}
\end{minipage}
\end{center}


Then we consider the non-trivial as well as non-singular rational solution \eqref{solu-2}, which is depicted in Fig. 2.


\begin{center}
\begin{picture}(120,70)
\put(-180,-23){\resizebox{!}{3cm}{\includegraphics{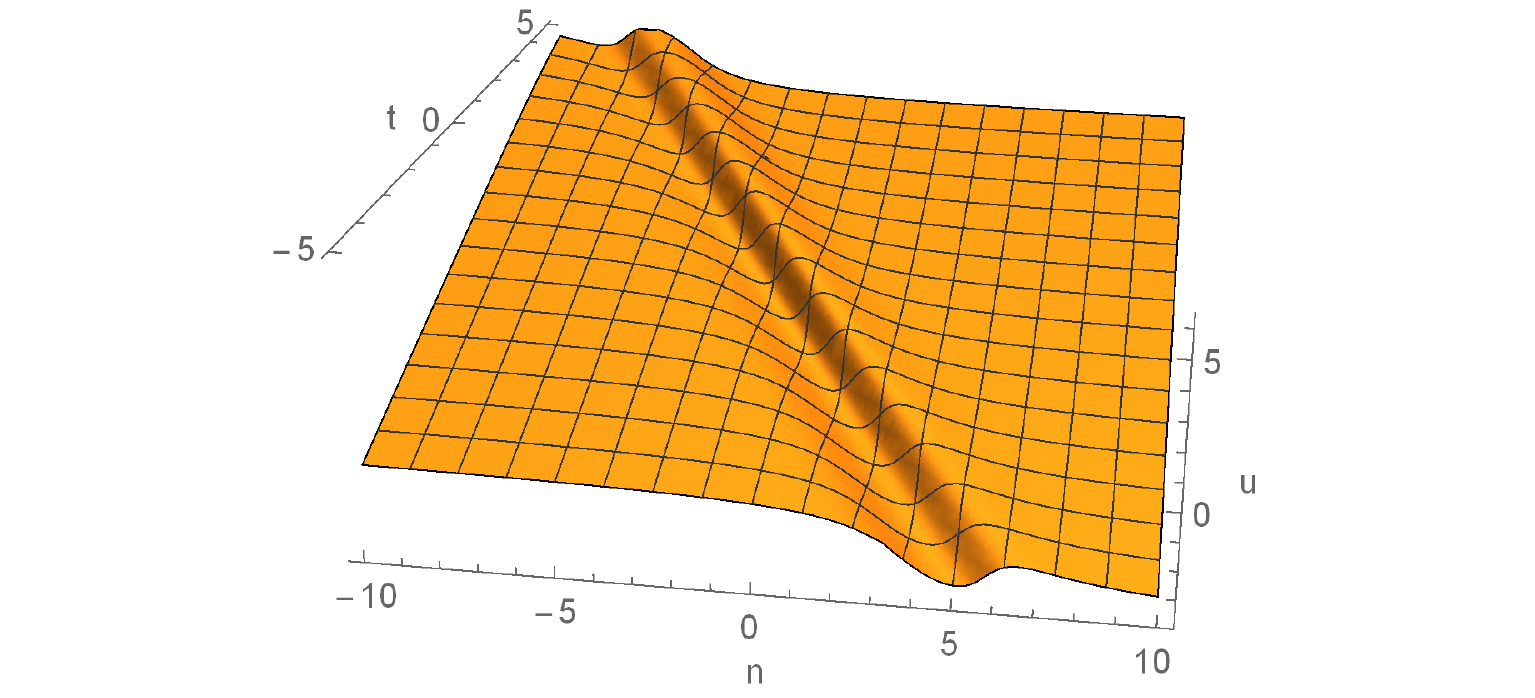}}}
\put(20,-23){\resizebox{!}{3cm}{\includegraphics{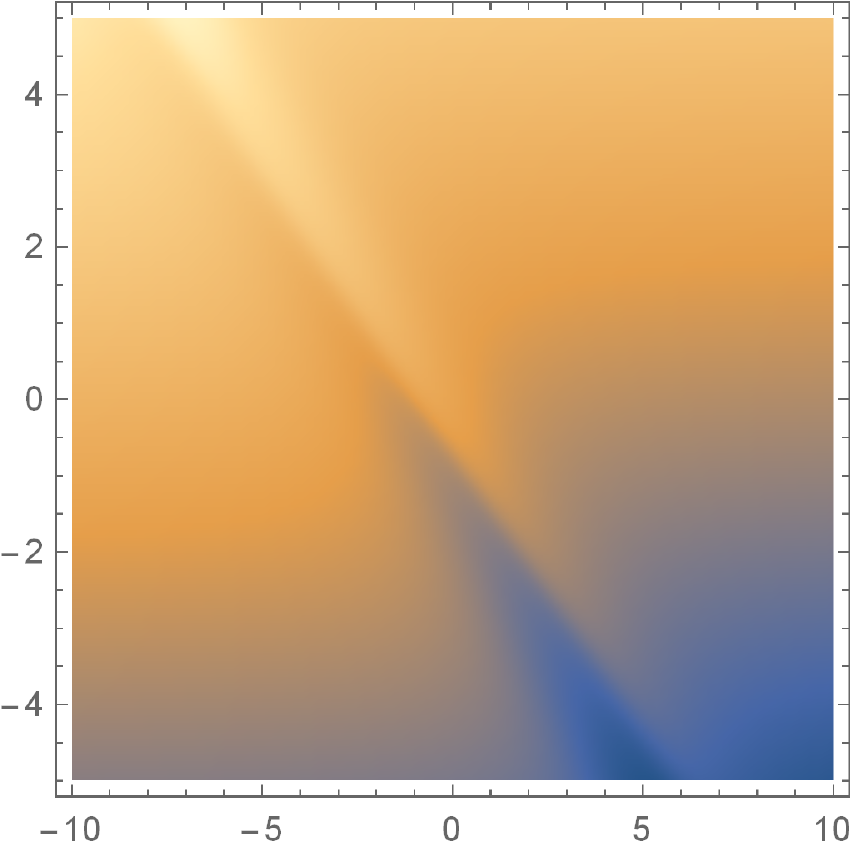}}}
\put(150,-23){\resizebox{!}{3cm}{\includegraphics{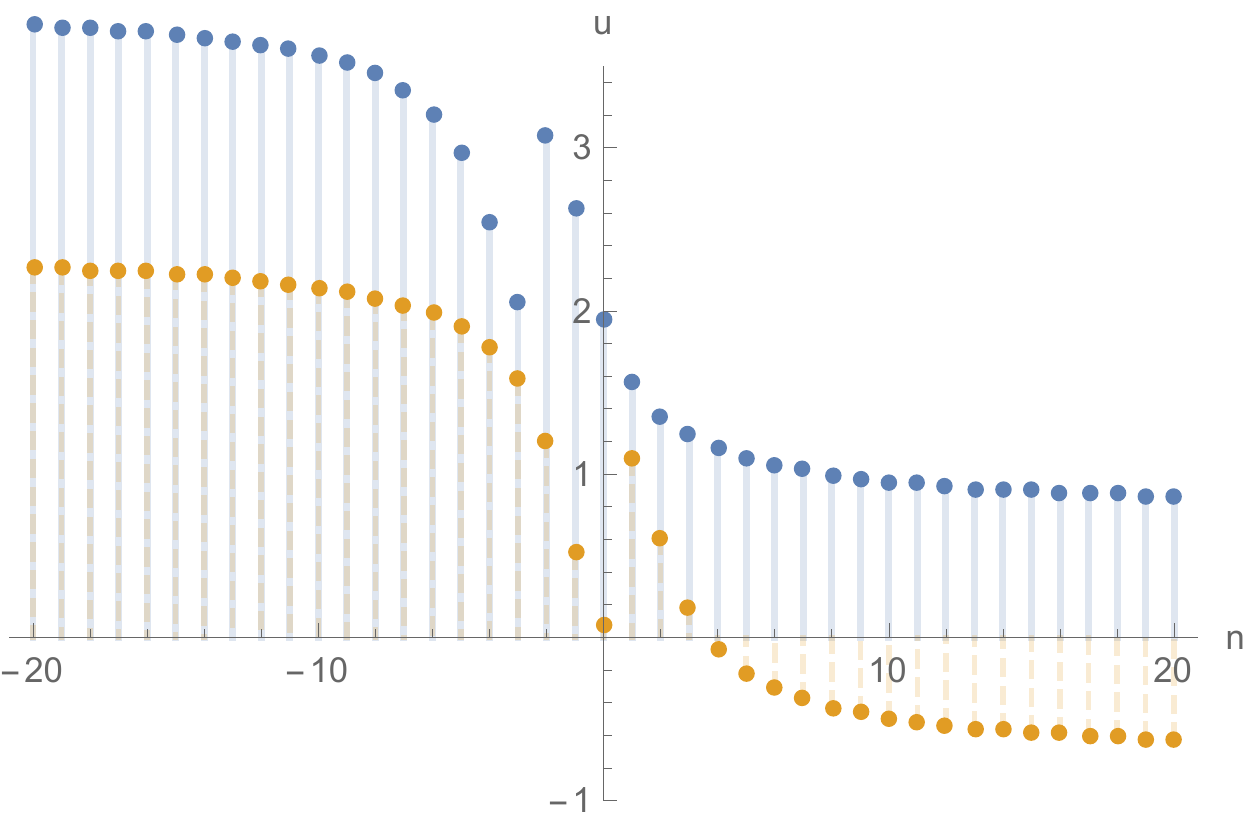}}}
\end{picture}
\end{center}
\vskip 20pt
\begin{center}
\begin{minipage}{11cm}{\footnotesize
(a)\qquad\qquad\qquad\qquad\qquad\qquad\qquad\qquad (b) \qquad\qquad\qquad\qquad\qquad\qquad (c)\\
{\bf Fig. 2} Rational solution given by \eqref{solu-2} for $c=1.2$. (a) Shape and movement. (b)
Density plot of (a) with larger range $n\in[-10, 10], t\in [-5,5]$.
The bright areas indicate the value of $u$ is positive and dark areas negative;
(c) Waves in blue and yellow stand for $t=1, -1$, respectively.}
\end{minipage}
\end{center}


In order to realize the asymptotic behavior analytically, we put this solution
in the following coordinates system
\begin{align}
(y=(1+z+n)c,z),
\end{align}
which yields
\begin{align}
\label{solu-2-yz}
u=\frac{\pi}{2}+\frac{c}{2}z+\arctan\bigg(\frac{8y+zc^3}
{6(1-1/4c^2+y^2)}-\frac{y}{3}\bigg).
\end{align}
It can be noticed that for a fixed $z$, $u\rightarrow \frac{c}{2}z$ for $y\rightarrow +\infty$ and
$u\rightarrow \pi+\frac{c}{2}z$ for $y\rightarrow -\infty$.

\section{Rational solutions for the sd-lpmKdV equation \eqref{sdpmKdV}}

In this section, rational solutions of the sd-lpmKdV equation \eqref{sdpmKdV}
will be constructed. We start by presenting some transformations of the equation
\eqref{sdpmKdV}, which makes it more managable. Introducing the variable $V_n \doteq V(n,t)$ by
\begin{align}
\label{v-V}
v_n=\left(\frac{q+c}{q-c}\right)^{\frac{n}{2}} e^{\frac{-ct}{q^2-c^2}} V_n,
\end{align}
and setting the parameter $p$ as
\begin{align}
\label{p-q}
p=\frac{q^2-c^2}{q},
\end{align}
we rearrange the equation \eqref{sdpmKdV} as
\begin{align}
\label{sdpmKdV-c}
\frac{\partial\ln V_{n}}{\partial t}=\frac{V_{n-1}-V_{n+1}}{(q-c)V_{n-1}+(q+c)V_{n+1}},
\end{align}
where $c$ is a non-zero constant.

\subsection{Bilinear equations and Casorati determinant solutions}

Through the dependent transformation
\begin{align}
\label{V-Tran}
V_n=\frac{g_n}{f_n},
\end{align}
the equation \eqref{sdpmKdV-c} is transformed into the bilinear equations
\begin{subequations}
\label{bili-1}
\begin{align}
& \label{bili-1a} (q D_t+\sinh D_n)g_n\cdot f_n=0, \\
& \label{bili-1b} (\cosh D_n+\frac{c}{q}\sinh D_n-1)g_n\cdot f_n=0.
\end{align}
\end{subequations}

Casorati determinant solutions to the bilinear system \eqref{bili-1} can be
summarized in the following Theorem from the analogue in our recent paper \cite{ZZ-ABS-rs},
and the following results can be verified by direct calculation.
\begin{Thm}
The Casorati determinants
\begin{align}
f_n=|\wh{N-1}|, \quad g_n=|-1,\wh{N-2}|
\end{align}
composed by $\phi(n,l) \doteq \phi(n,t,l)=(\phi_1(l),\phi_2(l),\cdots,\phi_N(l))^T$ solve the bilinear system \eqref{bili-1},
provided the basic column vectors $\phi(n,l)$ satisfy the CES
\begin{subequations}
\label{sdpmKdV-CES}
\begin{align}
& (q-c)\phi(n,l)=\phi(n+1,l)-\phi(n,l+1), \\
& (q+c)\varphi(n+1,l)=\varphi(n,l)+\varphi(n+1,l+1), \\
& \phi(n,l)=C(n)\varphi(n,l),\\
& C(n+1)\partial_t\phi(n,l)=C(n)(q\phi(n,l)-\phi(n+1,l)),
\end{align}
\end{subequations}
in which $\varphi(n,l)\doteq \varphi(n,t,l)=(\varphi_1(l),~\varphi_2(l),\ldots,\varphi_N(l))^T$ is an auxiliary vector
and $C(n)$ is an $N\times N$ matrix only depends on $n$ while independent of $l$ and $t$.
\end{Thm}

\subsection{Solutions to CES \eqref{sdpmKdV-CES}}

Applying a similar solving procedure to the CES \eqref{CES}, we construct exact solutions of \eqref{sdpmKdV-CES} by considering the eigenvalue structure of
canonical forms of $C(n)$, which lead to soliton solutions, Jordan-block solutions and
rational solutions for the equation \eqref{sdpmKdV-c}. For our convenience, we denote $\phi(n,l)$ by $\phi(l)$ and $\varphi(n,l)$ by $\varphi(l)$ hereafter.

\subsubsection{Soliton solutions}

When  $C(n)$ is a diagonal matrix
\begin{align}
C(n)=\mbox{diag}\left((q^2-k^2_1)^n,~(q^2-k^2_2)^n,~\ldots,~(q^2-k^2_N)^n\right),
\end{align}
with distinct $\{k_j\}$, the elements of $\phi(l)$ can be taken as
\begin{align}
\phi_j=a^+_j(q-k_j)^n(c-k_j)^l e^{\frac{k_j t}{q^2-k_j^2}}+a^-_j(q+k_j)^n(c+k_j)^l e^{\frac{-k_j t}{q^2-k_j^2}},
\end{align}
and the auxiliary $\varphi(l)$ is composed by
\begin{align}
\varphi_j=a^+_j(q+k_j)^{-n}(c-k_j)^l e^{\frac{k_j t}{q^2-k_j^2}}+a^-_j(q-k_j)^{-n}(c+k_j)^l e^{\frac{-k_j t}{q^2-k_j^2}},
\end{align}
where $a^{\pm}_j \in \mathbb{R},(j=1,\cdots,N)$. This case leads to $N$-soliton solutions of \eqref{sdpmKdV-c}.

\subsubsection{Jordan-block solutions}

When $C(n)$ is a LTT matrix with entries $\gamma_{s,j}$,
\begin{align*}
C(n)=(\gamma_{s,j})_{N\times N},~~\gamma_{s,j}=\Biggl\{
\begin{array}{ll}
\frac{1}{(s-j)!}\partial^{s-j}_k(q^2-k^2)^n,&~s\geq j,\\
0,&~s<j,
\end{array}
\end{align*}
the vectors $\phi(l)$ which are used to generate Casorati determinant can be taken as
\begin{subequations}
\begin{align}
\phi(l)&=\mathcal{B}^+\phi^+(l)+\mathcal{B}^-\phi^-(l)
\end{align}
with
\begin{align}
\phi^{\pm}(l)&=(\phi^{\pm}_0(l),\phi^{\pm}_1(l),\ldots,\phi^{\pm}_{N-1}(l))^T, \\
\phi^\pm_s(l)&=\frac{1}{s!}\partial_{k}^{s}[a^\pm (q\mp k)^n(c\mp k)^l e^{\frac{\pm k t}{q^2-k^2}}], \label{a+-}
\end{align}
\end{subequations}
where $a^{\pm} \in \mathbb{R}$ and $\mathcal{B}^{\pm}$ are two arbitrary
LTT matrices of $N$-th order. This case leads to Jordan-block solutions of \eqref{sdpmKdV-c}.

\subsubsection{Rational solutions}

Similar to the discussion given in Sec. \ref{subsub-rs-NLSNE}, to derive the non-trivial rational solutions
we now introduce the generating function
\begin{align}
\sigma(l)=\phi^+_0(l)+\phi^-_0(l)=(q-k)^{n}(c-k)^l e^{\frac{k t}{q^2-k^2}}+(q+k)^n(c+k)^l e^{\frac{-kt}{q^2-k^2}},
\end{align}
where we have chosen $a^{\pm}=1$ in \eqref{a+-}. The Taylor
expansion of $\sigma(l)$ with respect to $k$ at $k=0$ is written as
\begin{align}
\sigma(l)=\sum_{j=1}^{+\infty}\chi_{j+1}k^{2j},
\end{align}
where
\begin{align}
\chi_{j+1}=\frac{1}{(2j)!}\frac{\partial^{2j}}{\partial k^{2j}}\sigma(l)\bigg|_{k=0}, \quad (j= 0, 1, \ldots).
\end{align}
Then we get the vector $\chi=(\chi_1,\chi_2,\ldots,\chi_N)^T$ satisfying the CES \eqref{sdpmKdV-CES} with
\begin{align*}
C(n)=(\gamma^0_{s,j})_{N\times N},~~\gamma^0_{s,j}=\Biggl\{
\begin{array}{ll}
\frac{1}{(2(s-j))!}\partial^{2(s-j)}_k(q^2-k^2)^n\bigg|_{k=0},&~s\geq j,\\
0,&~s<j.
\end{array}
\end{align*}
Thus
\begin{align}
\label{fg-chi}
f=\mbox{Cas}(\chi)=|\wh{N-1}|, \quad  g=\mbox{Cas}(\chi)=|-1,\wh{N-2}|,
\end{align}
also solve the bilinear equations \eqref{sdpmKdV-CES} and
provide rational solutions for the equation \eqref{sdpmKdV-c}.

\subsection{Rational solutions to sd-lpmKdV equation \eqref{sdpmKdV} and dynamics}

By using the transformation \eqref{v-V}, one derives the rational solutions for the sd-lpmKdV equation \eqref{sdpmKdV} with \eqref{p-q}
from \eqref{fg-chi} immediately. Here we write out the first two non-trivial explicit forms for $v$
\begin{subequations}
\label{solu-v}
\begin{align}
& \label{solu-v1}
v=\frac{1}{c^2}\left(\frac{q+c}{q-c}\right)^{\frac{n}{2}} e^{\frac{-ct}{q^2-c^2}}\left(\frac{q^2}{c(t-nq)}+1\right), \\
&
v=\frac{1}{c^3}\left(\frac{q+c}{q-c}\right)^{\frac{n}{2}} e^{\frac{-ct}{q^2-c^2}}\left(\frac{3q^2(t-nq)(c(t-nq)+q^2)}{c^2\left((t-nq)^3-q^2(3t-nq)\right)}+1\right),
\end{align}
\end{subequations}
where we have taken $l=0$.

The solutions \eqref{solu-v} have an exponential growing
background part $\frac{1}{c^N}\left(\frac{q+c}{q-c}\right)^{\frac{n}{2}} e^{\frac{-ct}{q^2-c^2}}$.
In order to show the behaviors of rational solutions we ignore this part and only discuss
the part $\nu$, arising from $v=\frac{1}{c^N}\left(\frac{q+c}{q-c}\right)^{\frac{n}{2}} e^{\frac{-ct}{q^2-c^2}}(\nu+1)$.
It is easy to see that there is a singularity of the function
\begin{align}\label{nu-1}
\nu=\frac{q^2}{c(t-nq)}.
\end{align}
Obviously, the singular points are on a straight line
$n(t)=\frac{t}{q}$. We depict this function in Fig. 3.


\begin{center}
\begin{picture}(120,90)
\put(-100,-23){\resizebox{!}{4cm}{\includegraphics{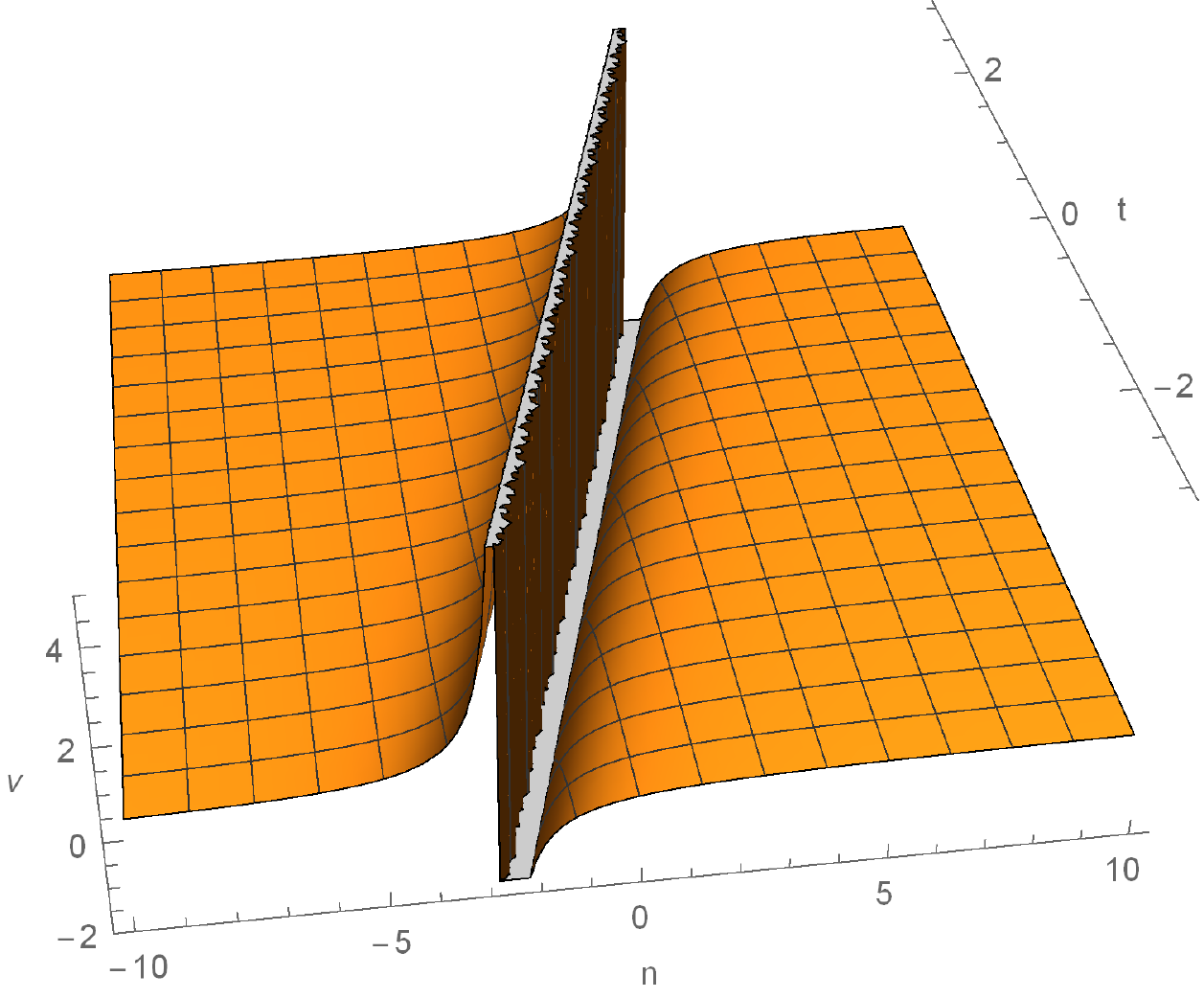}}}
\put(70,-23){\resizebox{!}{4cm}{\includegraphics{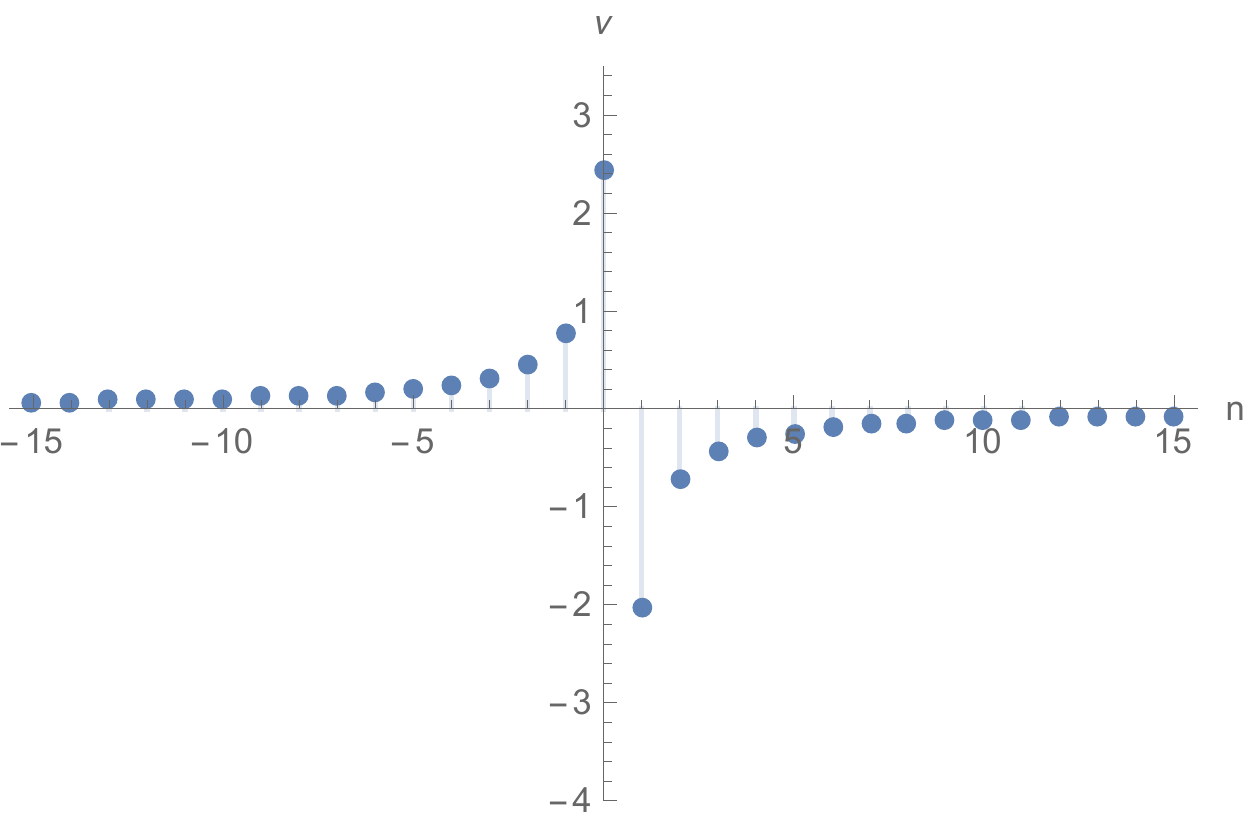}}}
\end{picture}
\end{center}
\vskip 20pt
\begin{center}
\begin{minipage}{11cm}{\footnotesize
\qquad\qquad\qquad(a)\qquad\qquad\qquad\qquad\qquad\qquad\qquad\qquad\qquad (b)\\
{\bf Fig. 3} The shape and movement of function \eqref{nu-1} for $q=1.1$ and $c=1$. (a) 3D-plot ; (b)
2D-plot for $t=0.5$.}
\end{minipage}
\end{center}


The function
\begin{align}\label{nu-2}
\nu=\frac{3q^2(t-nq)(c(t-nq)+q^2)}{c^2\left((t-nq)^3-q^2(3t-nq)\right)},
\end{align}
is also singular along the point trace
\[n(t)=\frac{t}{q}+\frac{1}{\sqrt{3}}\left[\left(\frac{\sqrt{27t^2-q^2}-3\sqrt{3}t}{q}\right)^{\frac{1}{3}}+
\left(\frac{q}{\sqrt{27t^2-q^2}-3\sqrt{3}t}\right)^{\frac{1}{3}}\right].\]
We depict the shape of this function in Fig. 4.


\begin{center}
\begin{picture}(120,90)
\put(-200,-23){\resizebox{!}{4cm}{\includegraphics{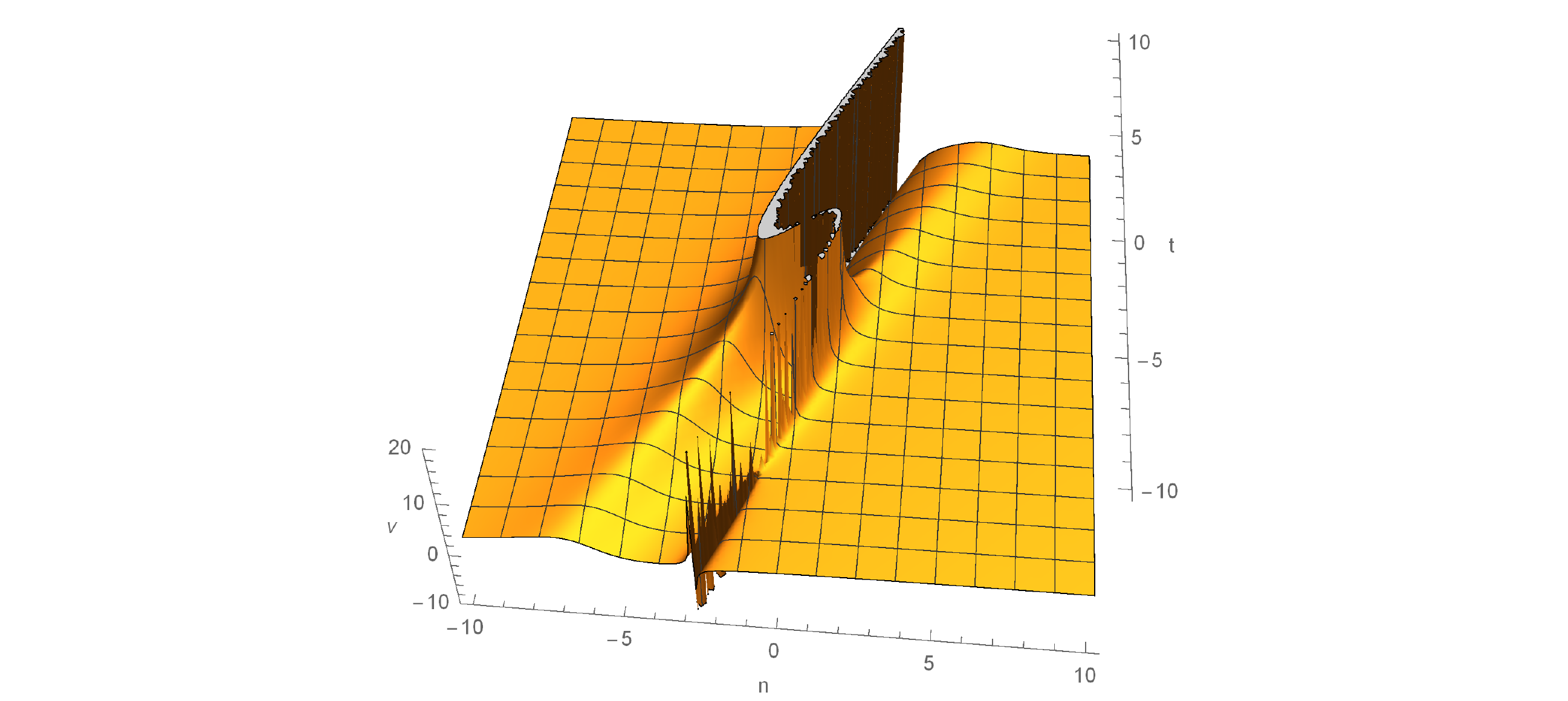}}}
\put(20,-23){\resizebox{!}{3cm}{\includegraphics{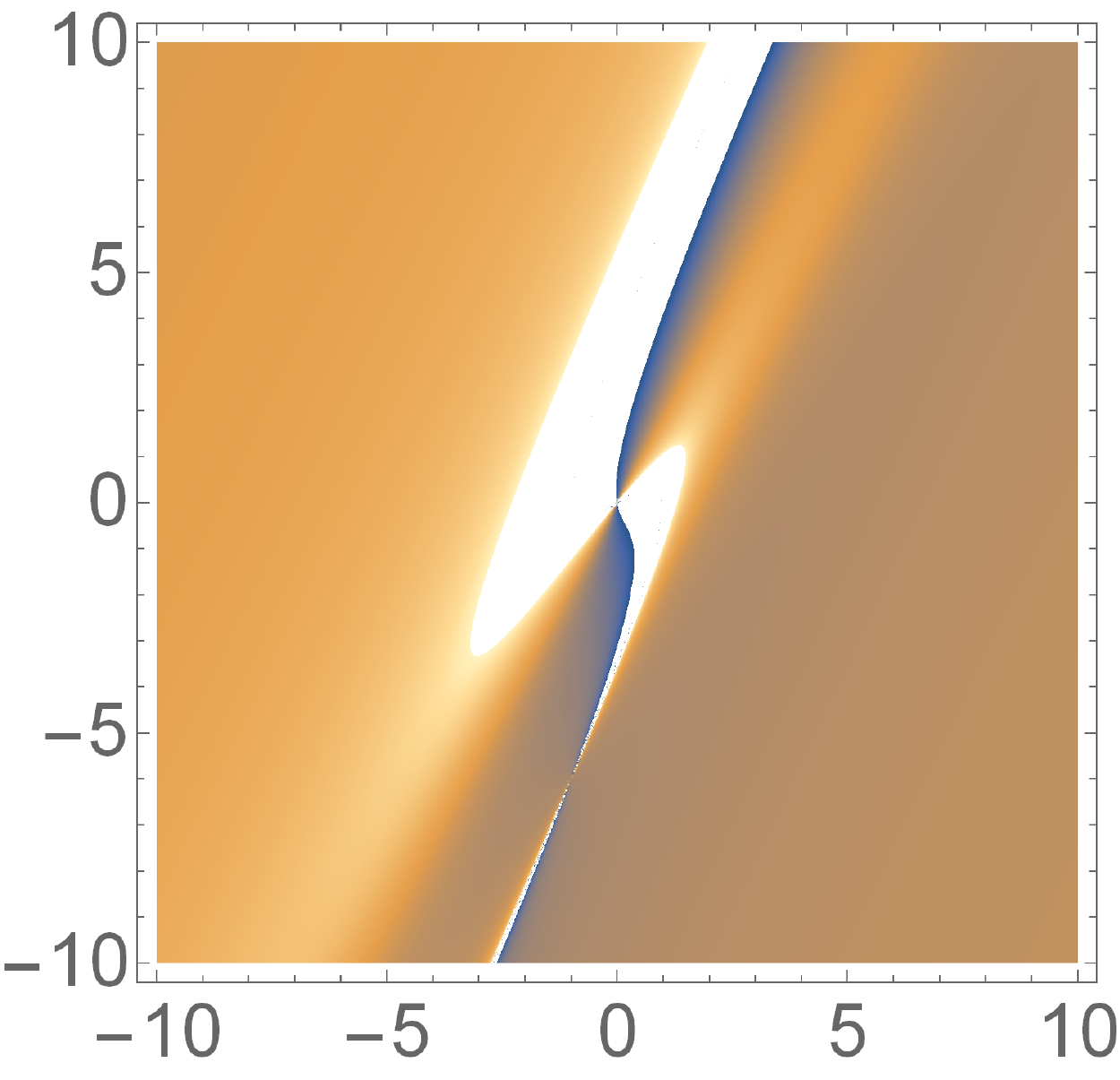}}}
\put(150,-23){\resizebox{!}{3cm}{\includegraphics{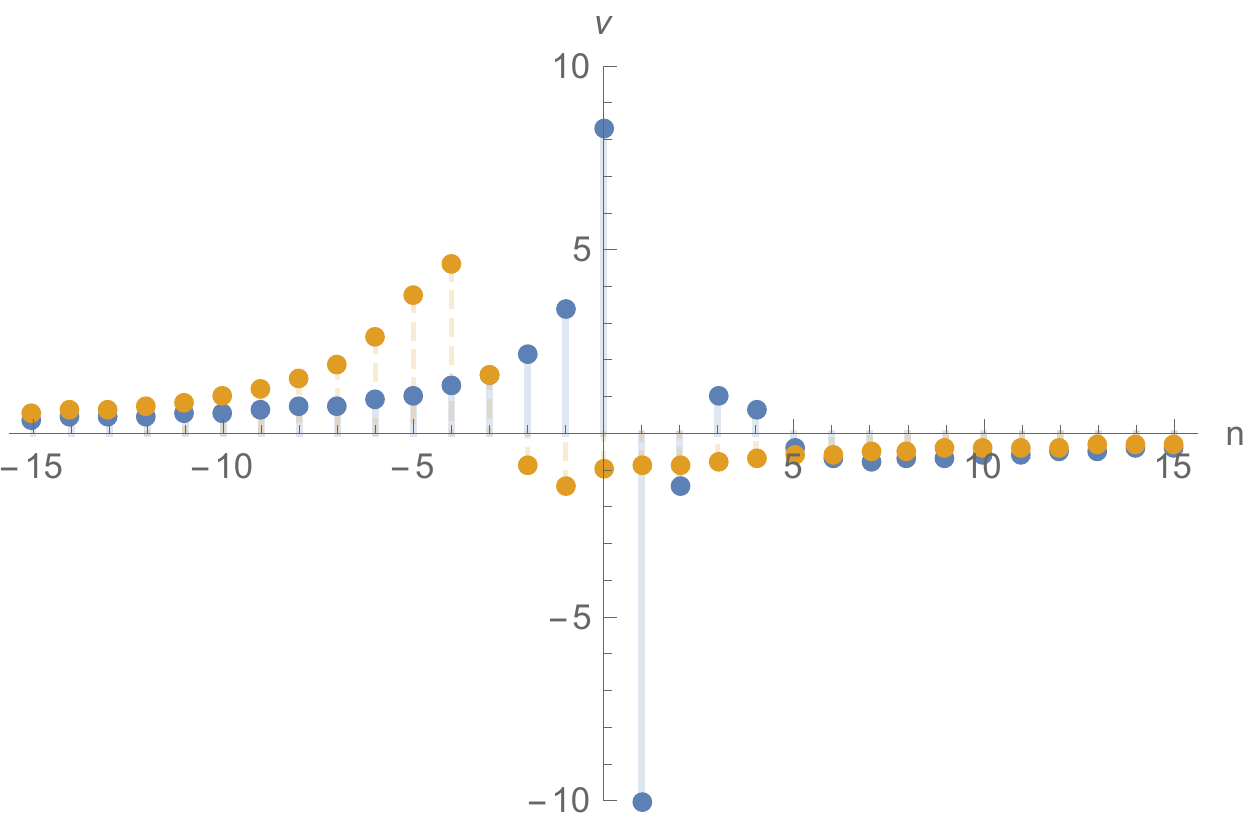}}}
\end{picture}
\end{center}
\vskip 20pt
\begin{center}
\begin{minipage}{11cm}{\footnotesize
(a)\qquad\qquad\qquad\qquad\qquad\qquad\qquad\qquad (b) \qquad\qquad\qquad\qquad\qquad\qquad \quad (c)\\
{\bf Fig. 4} Function given by \eqref{nu-2} for $q=2$ and $c=1$. (a) Shape and movement. (b)
Density plot of (a) with larger range $n\in[-10, 10], t\in [-10,10]$. Note that the white regions
indicate singularities.
(c) Waves in blue and yellow stand for $t=5, -5$, respectively.}
\end{minipage}
\end{center}


For the sake of realizing the asymptotic behavior analytically, we rewrite the function $\nu$ \eqref{nu-2} in terms of the following coordinates
\begin{align}
(\theta=t-nq,t),
\end{align}
which yields
\begin{align}
\nu=\frac{3q^2\theta(c\theta+q^2)}{c^2\left(\theta^3-q^2(2t+\theta)\right)}.
\end{align}
It can be noticed that for a fixed $t$, $\nu\rightarrow 0$ for $\theta\rightarrow \pm \infty$.

\section{Rational solutions for the sd-mKdV equation \eqref{sdmKdV} }

In this section, we will construct rational solutions of the sd-mKdV equation \eqref{sdmKdV}.
For this purpose, it is convenient to transform  the sd-mKdV equation \eqref{sdmKdV} into
\begin{align}
\label{sdmKdV-c}
\partial_z W_n=\left(1+(c^2+1)W^2_n-2c W_n\right)(W_{n-1}-W_{n+1}),
\end{align}
by a Galilean transformation
\begin{align}
\label{tran}
w(n,t)=(c^2+1)W(n,z)-c, \quad t=\frac{1}{c^2+1}z,
\end{align}
where $c$ is a non-zero constant.

\subsection{Bilinearisation of \eqref{sdmKdV-c} and Casorati determinant solutions}

By introducing logarithmic transformation
\begin{align}
\label{v-Tran}
W_n=\frac{i}{2}\ln\frac{f^*_n}{f_n},
\end{align}
the equation \eqref{sdmKdV-c} can be transformed into the bilinear equations
\begin{subequations}
\label{bili-3}
\begin{align}
& \label{bili-3a} (D_z+2\sinh D_n)f_n\cdot f^*_n=0, \\
& \label{bili-3b} (-ic D_z+2\cosh D_n-2)f_n\cdot f^*_n=0.
\end{align}
\end{subequations}

Exact Casorati determinant solutions to bilinear system \eqref{bili-3} can be summarized in the following Theorem.
\begin{Thm}
\label{Thm-CES-3}
The Casorati determinant
\begin{align}
\label{f-phi}
f_n=|\wt{N-1}|
\end{align}
solves the bilinear equations \eqref{bili-3},
provided the basic column vectors $\phi(n)\doteq \phi(n,z)=(\phi_1(n),\phi_2(n),\cdots,\phi_N(n))^T$ satisfy
the following CES
\begin{subequations}
\label{CES-3}
\begin{align}
& \phi(n+1)+\phi(n-1)=A\phi(n), \\
& i(\phi(n-1)-\phi(n+1))+cA\phi(n)=B\phi^*(n), \\
& \partial_z \phi(n)=\phi(n-2).
\end{align}
\end{subequations}
Here, $A=(a_{ij})_{N\times N}$ and $B=(b_{ij})_{N\times N}$
are two $N\times N$ invertible matrices and satisfy
\begin{align}
\label{AB-sdmKdV}
B^2=(c^2+1)A^2-4I.
\end{align}
\end{Thm}

\subsection{Solutions for CES \eqref{CES-3}}

Analogous to the previous analysis, in this part we list several choices
of matrices $A$ and $B$, and present the corresponding solutions to the CES \eqref{CES-3}.

\subsubsection{Soliton solutions}

When $A$ is a diagonal matrix defined as \eqref{A1}, from the relation \eqref{AB-sdmKdV} we have
\begin{align} \label{B3}
B=\mbox{diag}\left(2\sqrt{c^2\cosh^2k_1+\sinh^2k_1},~\ldots,~2\sqrt{c^2\cosh^2k_N+\sinh^2k_N}\right).
\end{align}
In this case, the explicit form of Casorati elements $\phi_j$ is
\begin{align}
& \phi_j(n)=a^+_j\sqrt{c(e^{k_j}+e^{-k_j})+(e^{k_j}-e^{-k_j})i}e^{k_j n+e^{-2 k_j}z} \nn \\
&\qquad\qquad +a^-_j\sqrt{c(e^{k_j}+e^{-k_j})-(e^{k_j}-e^{-k_j})i}e^{-k_j n+e^{2 k_j}z},
\end{align}
where $k_j, a^{\pm}_j \in \mathbb{R}, ~(j=1,\cdots,N)$. This yields the $N$-soliton solutions for the equation \eqref{sdmKdV-c}.

\subsubsection{Jordan-block solutions}

When $A$ and $B$ are defined as LTT matrices, written as
\begin{subequations}
\label{AB-J3}
\begin{align}
& A=(\mu_{s,j})_{N\times N},\quad \mu_{s,j}=\Biggl\{
\begin{array}{ll}
\frac{2}{(s-j)!}\partial^{s-j}_k\cosh k,&~s\geq j,\\
0,&~s<j,
\end{array}\\
& B=(\nu_{s,j})_{N\times N},\quad \nu_{s,j}=\Biggl\{
\begin{array}{ll}
\frac{1}{(s-j)!}\partial^{s-j}_k\sqrt{4c^2\cosh^2k+4\sinh^2k},&~s\geq j,\\
0,&~s<j,
\end{array}
\end{align}
\end{subequations}
which satisfy the relation \eqref{AB-sdmKdV}.
Then the vectors $\phi(n)$ which are used to generate the Casorati determinant can be taken as
\begin{subequations}
\label{Ph-lim-Ph-3}
\begin{align}
\label{Ph-lim-3}
\phi(n)&=\mathcal{C}^+\phi^+(n)+\mathcal{C}^-\phi^-(n)
\end{align}
with
\begin{align}
\label{Ph-i-3}
\phi^{\pm}(n)&=(\phi^{\pm}_0(n),\phi^{\pm}_1(n),\ldots,\phi^{\pm}_{N-1}(n))^T, \\
\label{phi-k-3}
\phi^\pm_s(n)&=\frac{1}{s!}\partial_{k}^{s}[a^{\pm}\sqrt{c(e^{k}+e^{-k})+(e^{k}-e^{-k})i}e^{\pm k n+e^{\mp 2k}z}],
\end{align}
\end{subequations}
where $a^{\pm} \in \mathbb{R}$ and $\mathcal{C}^{\pm}$ are two arbitrary LTT matrices of $N$-th order.

\subsubsection{Rational solutions}

To proceed, we take $a^{\pm}=1$ in \eqref{phi-k-3} and consider function
$\varrho(n)=\phi^+_0(n)+\phi^-_0(n)$, i.e.,
\begin{align}
\label{varrho-ra}
\varrho(n)=\sqrt{c(e^{k}+e^{-k})+(e^{k}-e^{-k})i}e^{k n+e^{-2k}z}+\sqrt{c(e^{k}+e^{-k})-(e^{k}-e^{-k})i}e^{-kn+e^{2k}z}.
\end{align}
The Taylor expansion of $\varrho(n)$ with respect to $k$ at $k=0$ can be written as
\begin{align}
\varrho(n)=\sum_{j=1}^{+\infty}\omega_{j+1}k^{2j},
\end{align}
where
\begin{align}
\omega_{j+1}=\frac{1}{(2j)!}\frac{\partial^{2j}}{\partial k^{2j}}\varrho(n)\bigg|_{k=0}, \quad (j= 0, 1, \ldots).
\end{align}
Then we obtain the vectors $\omega=(\omega_1,\omega_2,\ldots,\omega_N)^T$ satisfying the relations \eqref{CES} with
\begin{subequations}
\label{AB-J3-0}
\begin{align}
& A=(\mu^0_{s,j})_{N\times N},\quad \mu^0_{s,j}=\Biggl\{
\begin{array}{ll}
\frac{2}{(2(s-j))!},&~s\geq j,\\
0,&~s<j,
\end{array}\\
& B=(\nu^0_{s,j})_{N\times N},\quad \nu^0_{s,j}=\Biggl\{
\begin{array}{ll}
\frac{1}{(2(s-j))!}\partial^{2(s-j)}_k\sqrt{4c^2\cosh^2k+4\sinh^2k}\bigg|_{k=0},&~s\geq j,\\
0,&~s<j.
\end{array}
\end{align}
\end{subequations}
Moreover, $f_n$ composed by $\omega$ are solutions of the bilinear equation
\eqref{bili-3}. The first few explicit forms of some $f=\mbox{Cas}(\omega)$ are
\begin{subequations}
\begin{align}
& f_{N=1}=2 e^z \sqrt{2c}, \label{fn=31} \\
& f_{N=2}=8 e^{2z} (i+2(1+n-2z)c),  \label{fn=32}  \\
& f_{N=3}=\frac{16\sqrt{2} e^{3z}}{3c^{\frac{3}{2}}}\big(3i-6(2+n-2z)c+12i(2+n-2z)^2c^2+8\big((2+n-2z)^3-2-n+6z\big)c^3\big). \label{fn=33}
\end{align}
\end{subequations}

\subsection{Rational solutions for the sdmKdV equation \eqref{sdmKdV} and dynamics}

By the transformations \eqref{tran}, the rational solutions of the sdmKdV equation \eqref{sdmKdV} can be expressed as
\begin{align}
w=(c^2+1)\partial_z \arctan \frac{\mbox{Im}f_n}{\mbox{Re}f_n}-c,\quad z=(c^2+1)t.
\end{align}
By \eqref{fn=32}, we get the first non-trivial rational solution of \eqref{sdmKdV} reading
\begin{align}
\label{solu-1-sdmKdV}
w=-c+\frac{4c(1+c^2)}{1+4c^2(1+n-2(1+c^2)t)^2}.
\end{align}
Obviously, this solution is non-singular and provides a traveling wave which
propagates with a constant speed $2(1+c^2)$, as depicted in Fig. 5. For a
given $t$, the wave has a single extreme value $w=c(4c^2+3)$, which appears at point trace $n(t)=2(1+c^2)t-1$. In terms of the
sign of the parameter $c$, the ``amplitude'' of the wave can be positive or negative.


\begin{center}
\begin{picture}(120,90)
\put(-150,-23){\resizebox{!}{3.5cm}{\includegraphics{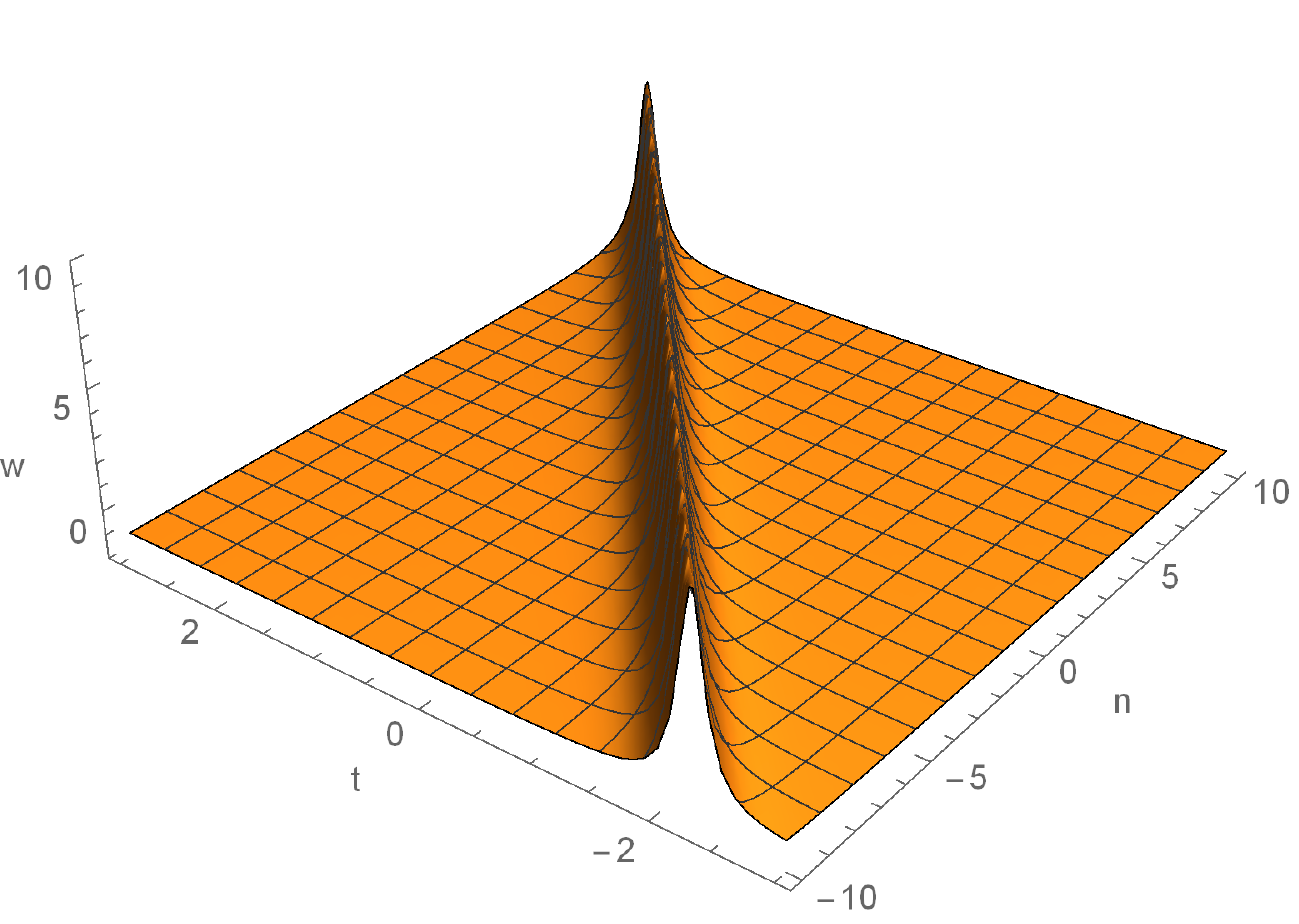}}}
\put(0,-23){\resizebox{!}{3cm}{\includegraphics{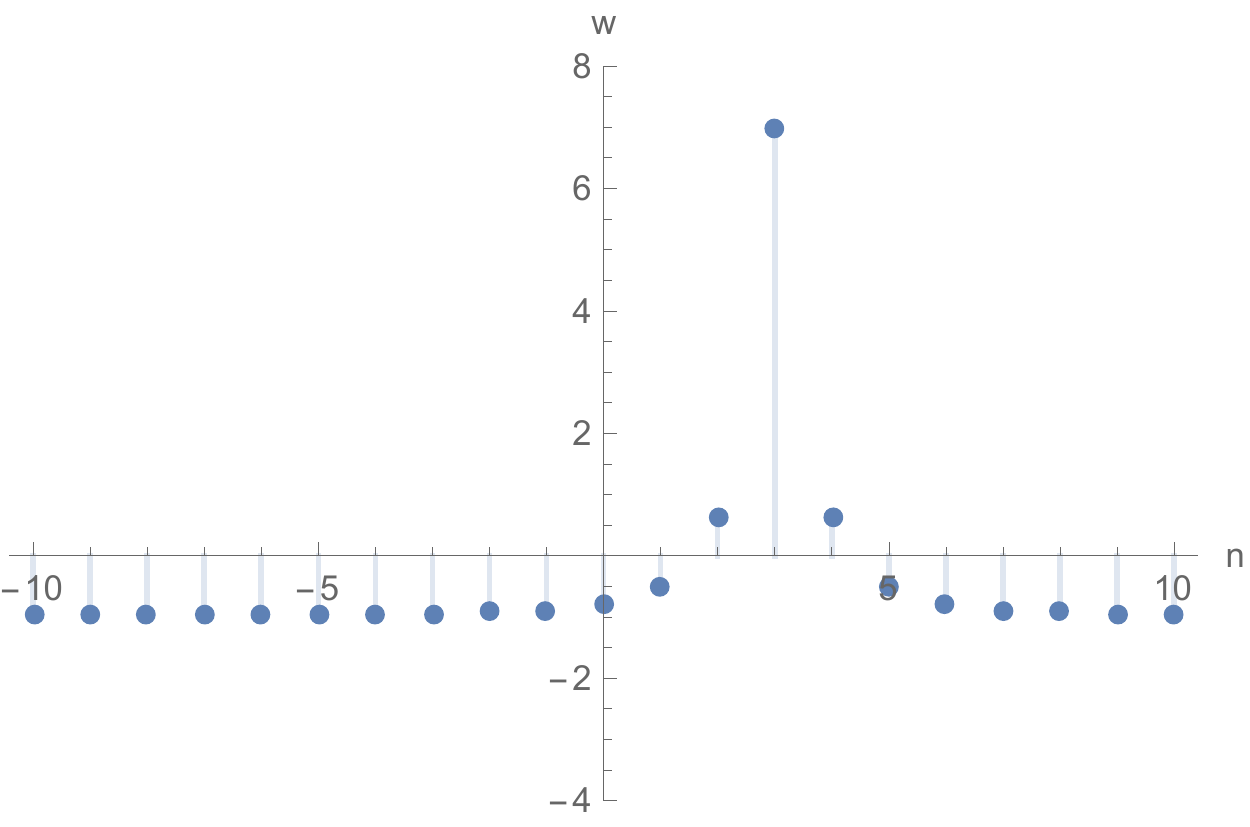}}}
\put(150,-23){\resizebox{!}{3cm}{\includegraphics{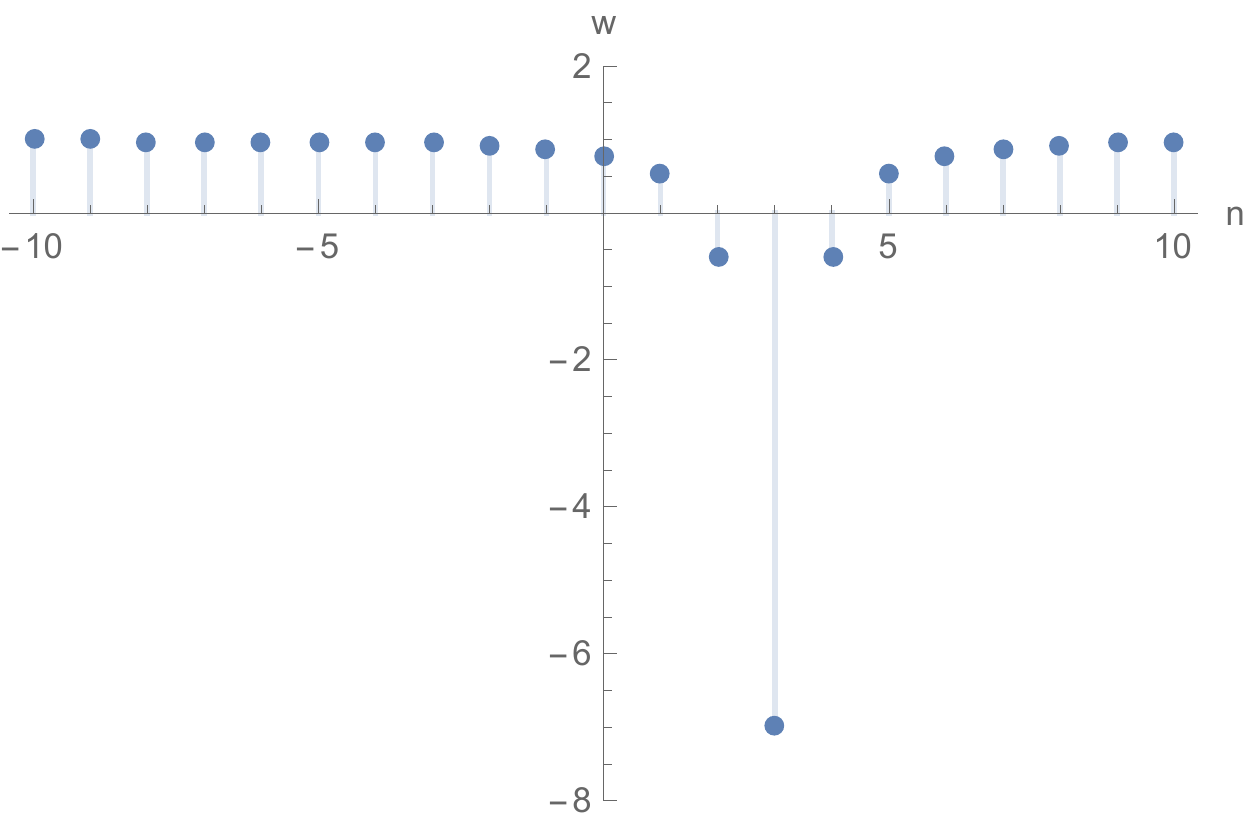}}}
\end{picture}
\end{center}
\vskip 20pt
\begin{center}
\begin{minipage}{11cm}{\footnotesize
\qquad(a)\qquad\qquad\qquad\qquad\qquad\qquad\qquad (b)\qquad\qquad\qquad\qquad\qquad \qquad\qquad(c) \\
{\bf Fig. 5} The shape and movement of solution \eqref{solu-1-sdmKdV} for $c=1$. (a) 3D-plot; (b)
2D-plot for $t=1$; (c) 2D-plot for $t=-1$.}
\end{minipage}
\end{center}


The rational solution related to \eqref{fn=33} reads
\begin{align}
\label{solu-2-sdmKdV}
w=-\frac{12c(c^2+1)(3+12c^2(1-2x^2)+16c^4x(8z-x^3))}{
9+108c^2x^2+48c^4x(-8z+2x+x^3)+64c^6(4z-x+x^3)^2}-c,
\end{align}
where $x=2+n-2z$ and $z=(c^2+1)t$. We depict the solution \eqref{solu-2-sdmKdV} in Fig. 6.
For a fixed $z$ and let $x\rightarrow \infty$, we find $w\rightarrow -c$.


\begin{center}
\begin{picture}(120,90)
\put(-150,-23){\resizebox{!}{3.5cm}{\includegraphics{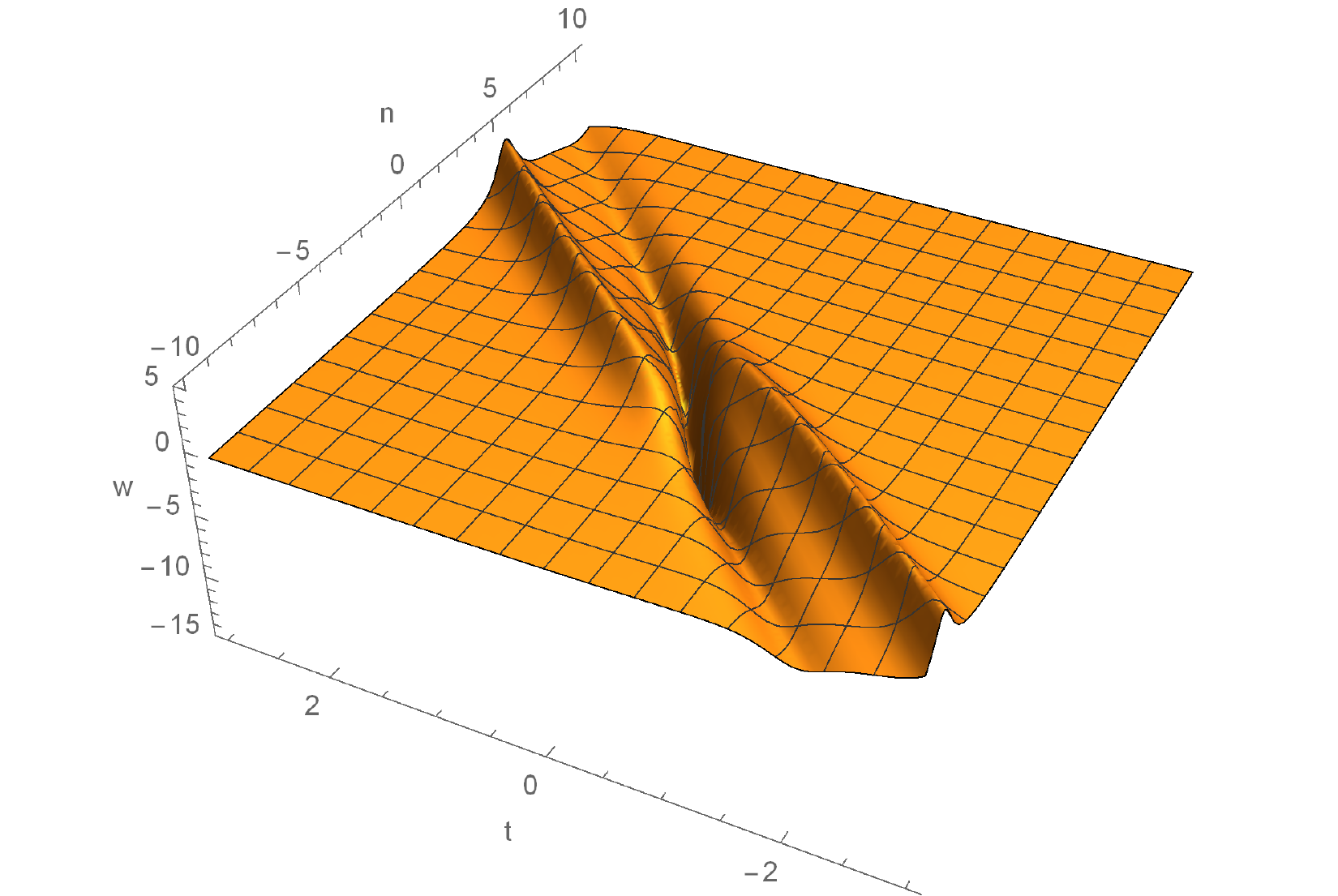}}}
\put(20,-23){\resizebox{!}{3cm}{\includegraphics{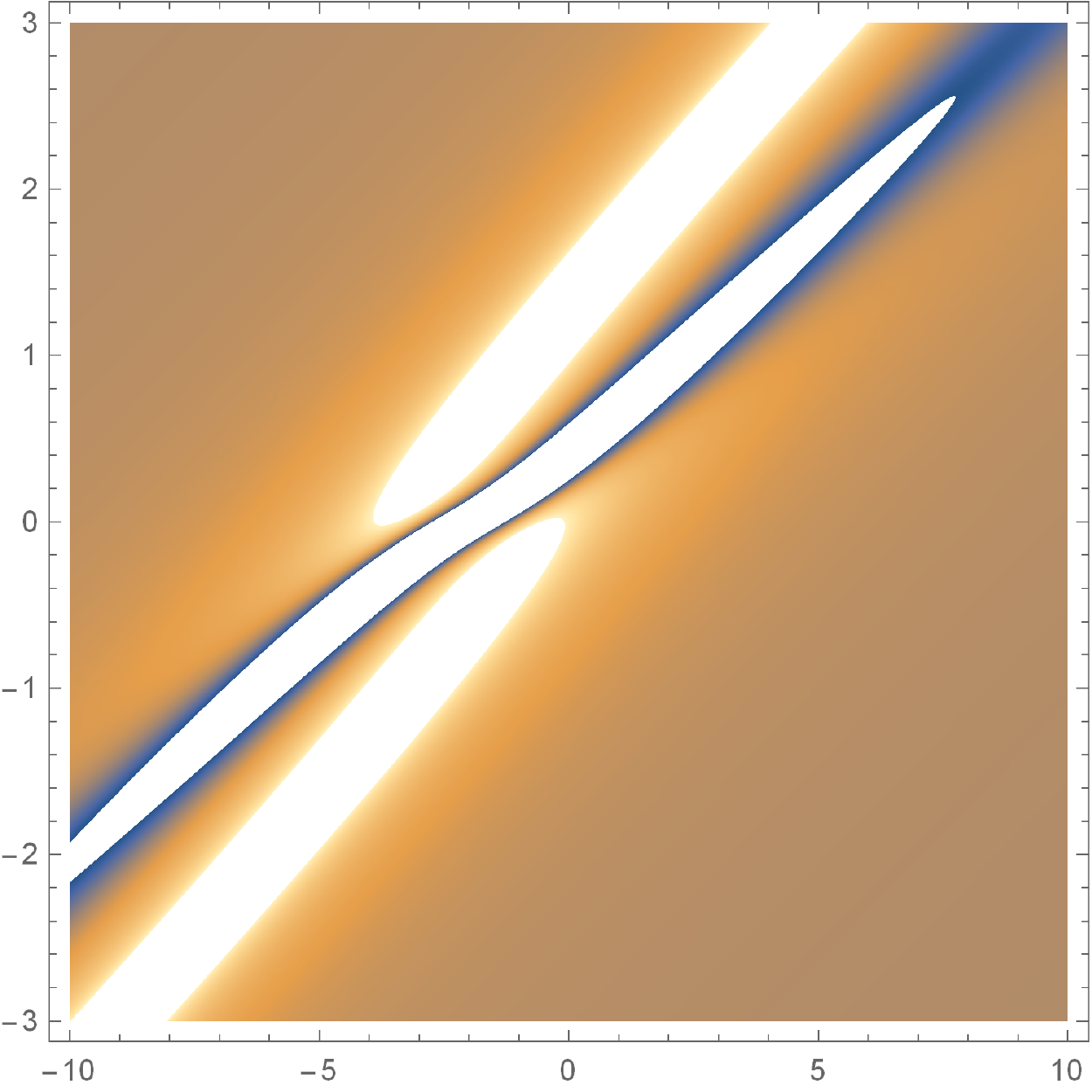}}}
\put(150,-23){\resizebox{!}{3cm}{\includegraphics{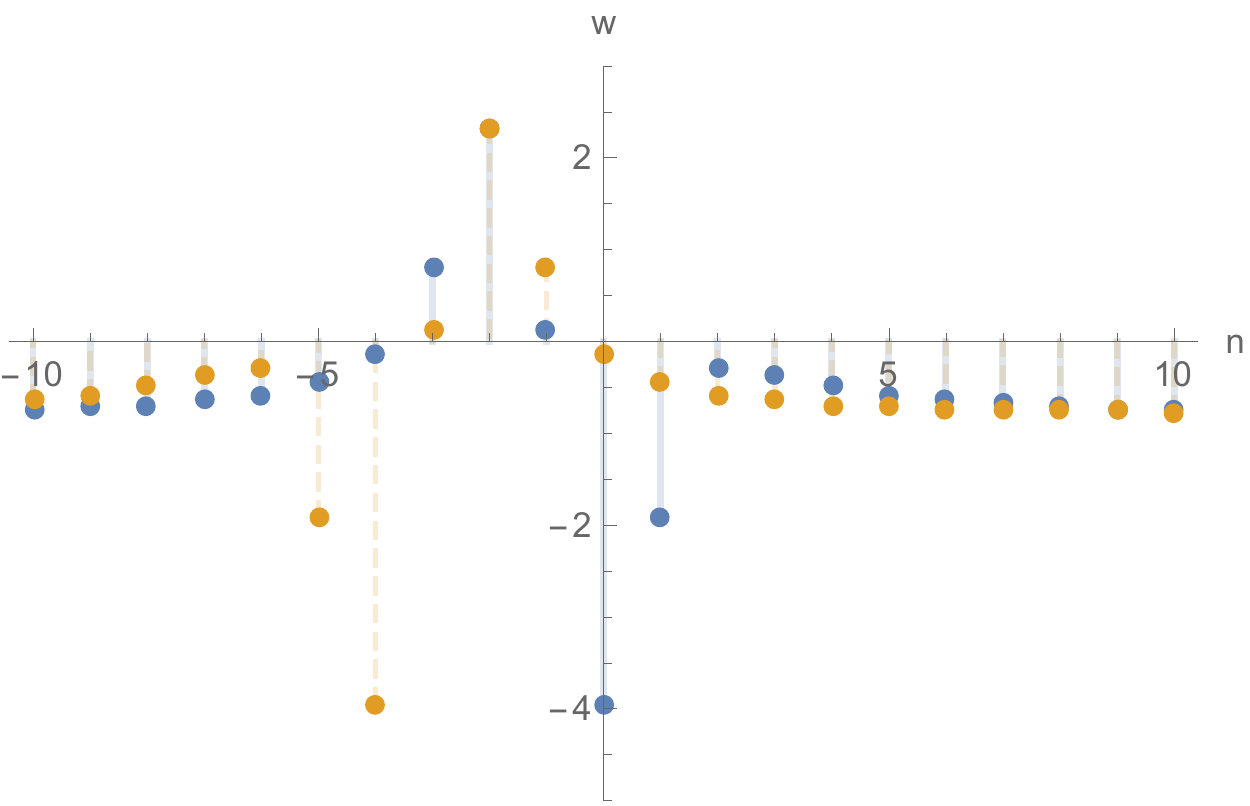}}}
\end{picture}
\end{center}
\vskip 20pt
\begin{center}
\begin{minipage}{11cm}{\footnotesize
\qquad(a)\qquad\qquad\qquad\qquad\qquad\qquad\qquad (b)\qquad\qquad\qquad\qquad\qquad \qquad\qquad(c) \\
{\bf Fig. 6} Solution given by \eqref{solu-2-sdmKdV} for $c=0.8$. (a) Shape and movement. (b)
Density plot of (a) with larger range $n\in[-10, 10], t\in [-3,3]$.
(c) Waves in blue and yellow stand for $t=0.5, -0.5$, respectively.}
\end{minipage}
\end{center}



\section{Conclusions}

The mKdV equation provides fruitful and interesting applications
in many physics and mathematics contexts. It is also important to study its semi-discrete forms.
There are many expressions for the semi-discrete mKdV equation due to different discretisations.
In this paper, we have dealt with three kinds of the semi-discrete mKdV equations.
The technique to construct Casorati determinant solutions in this paper was initially
motivated by the procedure used to derive the rational solutions for the mKdV equation.
Except for rational solutions of the equations \eqref{NLSNE}-\eqref{sdmKdV},
soliton solutions and Jordan blocks solutions are also presented.
Dynamics of some rational solutions are specially analysed and illustrated. In the future,
we would like to analyze exact solutions of other kinds of semi-discrete mKdV type equations.
In addition, we have made use of the connection between the NLSNE equation \eqref{NLSNE} and
the semi-discrete equation \eqref{sdmKdV} to obtain different rational solutions of the latter one (see Appendix \ref{app-1}).
It would be also interesting to clarify the relations between different forms of semi-discrete mKdV equations.

\vskip 20pt
\subsection*{Acknowledgments}
This project is supported by the Natural Science Foundation of Zhejiang Province (Nos.
LY17A010024, LY18A010033) and the National Natural Science Foundation of
China (No. 11401529).

\vskip 20pt

\begin{appendix}

\section{Rational solutions for the sd-mKdV equation \eqref{sdmKdV}} \label{app-1}

We now present the solutions of the equation \eqref{sdmKdV} by using the connection between the equation \eqref{NLSNE} and \eqref{sdmKdV} \cite{sdmkdv-1976}. 
We consider that $w_n$ in \eqref{sdmKdV} with even and odd $n$ have different asymptotic values at $\left| n \right| \rightarrow \infty$. Let
\begin{align}
s_n = w_{2n+1},~~r_n = w_{2n},
\end{align}
in \eqref{sdmKdV}, we have
\begin{subequations}
\begin{align}
& \label{odd-re} \partial_t s_n=(1+s_n^2)(r_{n}-r_{n+1}), \\
& \label{even-re1} \partial_t r_n=(1+r_n^2)(s_{n-1}-s_{n}).
\end{align}
\end{subequations}
The equation \eqref{even-re1} turns into
\begin{align}
\partial_t (\arctan r_n)=\partial_t (u_{n-1}- u_{n}),
\end{align}
by setting $s_n=\partial_t u_n$.
Then one immediately gets
\begin{align}
\label{re-u-r}
r_n=\tan  (u_{n-1}- u_{n}),
\end{align}
where we choose the integral constant be zero.
Substituting \eqref{re-u-r} into \eqref{odd-re} and making use of the relation
between $s_n$ and $u_n$, we obtain \eqref{NLSNE}. In this sense, solutions of the NLSNE \eqref{NLSNE}
obtained in subsection \ref{sec-212} can be employed to construct the related solutions of the sd-mKdV equation \eqref{sdmKdV}, i.e.,
\begin{align} \label{v-odd-even}
 w_{2n+1}=\partial_t u_n,\quad  w_{2n}=\tan  (u_{n-1}- u_{n}).
\end{align}

With regard to the rational solutions to the sd-mKdV equation \eqref{sdmKdV}, we can also construct them by
the above relation between $u_{n}$ and $v_{n}$. Here we list the explicit forms of several rational solutions of
the sd-mKdV equation \eqref{sdmKdV} derived from solutions \eqref{rational-NLSNE-2} of the NLSNE \eqref{NLSNE},
\begin{subequations}
\begin{align*}
& w_{2n+1} = \frac{c}{\sqrt{4-c^2}} \bigg(1- \frac{8}{4+(1+2n+\frac{4t}{\sqrt{4-c^2}})^2 c^2} \bigg), \\
& w_{2n} = \frac{4c}{4+c^2\big(2n+1+\frac{4t}{\sqrt{4-c^2}} \big)\big(2n-1+\frac{4t}{\sqrt{4-c^2}}\big)},
\end{align*}
and
\begin{align*}
& w_{2n+1} = \frac{c}{\sqrt{4-c^2}}- \frac{\partial_t h(n,t)}{1+h^2(n,t)}, \\
& w_{2n} = \frac{h(n,t)-h(n-1,t) }{1+h(n,t) h(n-1,t)} ,
\end{align*}
\end{subequations}
where $h(n,t)=\frac{c}{3}\big(1+\frac{2}{\sqrt{4-c^2}}t+n\big)-\frac{4\big(1+\frac{2}{\sqrt{4-c^2}}t+n\big)c+\frac{1}{\sqrt{4-c^2}}t c^3}
{3\big(1-1/4c^2+(1+\frac{2}{\sqrt{4-c^2}}t+n)^2c^2\big)}$.

\end{appendix}

\end{document}